\documentclass[twocolumn,showpacs,amsmath,amssymb,pra]{revtex4-1}

    \usepackage{graphicx}
    \usepackage{color}
    \usepackage{dcolumn}
    \usepackage{bm}
    \usepackage{xspace}
    \newcommand{\wf}{wavefunction\xspace}

    \newcommand{\IgnoreThis}[1]{#1}

\begin{document}

\title{Time Evolution of Harmonic Oscillator Thermal Momentum Superposition States}

\author{Ole Steuernagel}

\affiliation{School of Physics, Astronomy and Mathematics,
University of Hertfordshire, Hatfield, AL10 9AB, UK }


\email{O.Steuernagel@herts.ac.uk}

\date{\today}

\begin{abstract}
  The time evolution of thermal states of a mirror released from a
  tight harmonic trap is studied. After the release no dissipation is
  assumed to be present and the mirror is, after a time delay, kicked
  into a momentum superposition state. The thermal character of the
  initial state washes out the telltale interference patterns of the
  superposition but no loss of coherence is found. This investigation
  resolves a controversy about decoherence--without--dissipation and
  shows that  entrained measurements can be surprisingly insensitive
  to temperature effects.
\end{abstract}


\maketitle

\IgnoreThis{\section{Introduction}}

For massive objects it is hard to create a pure quantum state
unentangled with the environment. Many studies of the decohering
effects of the environment do however assume that the studied system
is in a pure state initially, unentangled with the environment.
Although realizable for atomic systems this is implausible for massive
objects such as opto-mechanical mirrors~\cite{Marquardt_Physics.2.40}.
Ford \emph{et~al.}~\cite{Ford_PRA01,Ford_PLA01,Ford_AJP02} are
therefore quite justified in insisting that for investigations of the
dynamics of quantum superposition states of massive systems, weakly
coupled to the environment, \emph{thermal initial} states should be
considered. They derived expressions for the loss of interference
contrast of formally thermalized momentum superpostion states of free
massive objects. Their interpretations have received a somewhat mixed
reception though. Notably, their interpretation of a process they
dubbed ``decoherence without dissipation"~\cite{Ford_PLA01} has been
either criticised~\cite{Gobert_PRA04,Ambegaokar_JSP06} or
ignored~\cite{Zurek_RMP03,Schlosshauer_RMP05}.

Here, following Ford \emph{et~al.'s} approach, a mirror is studied
which initially is weakly coupled to the environment such that we can
neglect dissipation other than assuming the mirror is in thermal
equilibrium with the laser beams that form a tight harmonic potential
holding and cooling it~\cite{Corbitt_PRL07}. When the lasers are
switched off the mirror becomes effectively decoupled from the
environment, we model this as a ballistically expanding mirror without
dissipation~\cite{Ford_PRA01,Ford_PLA01,Ford_AJP02}. Once the coherent
width of the mirror's spatial distribution is sufficient to imprint a
photon's momentum-recoil pattern the mirror is kicked into a momentum
superposition state and then investigated using the
`entrainment procedure' of reference~\cite{Ole_quantumEntrained}.

This work addresses two main questions. Can decoherence without
dissipation~\cite{Ford_PLA01} occur? How badly does the thermal
nature of the initial state affect the formation and persistence of
interference patterns of a mirror's wave
packet~\cite{Ole_quantumEntrained}?


Ford and O'Connell use the attenuation of the interference contrast
as a measure for decoherence. This is not the conventional approach
for the quantification of decoherence, which is based on the loss of
off-diagonal elements of the reduced density
matrix~\cite{Zurek_RMP03,Schlosshauer_RMP05,Gobert_PRA04,Ambegaokar_JSP06}.
One can therefore refute Ford and O'Connell's assertion, that they
found evidence for ``decoherence without
dissipation"~\cite{Ford_PLA01}, on formal grounds, arguing, that no
such process can occur without coupling to the
environment~\cite{Gobert_PRA04,Ambegaokar_JSP06}. I temporarily
adopt Ford and O'Connell's approach though and investigate the
visibility of the interference pattern, since it is the quantity an
experimentalist is most likely to
measure~\cite{Ford_PLA01,Ford_PRA01}.

Ford et~al.~\cite{Ford_PLA01,Ford_PRA01,Ford_AJP02} consider a free
particle in a Gaussian state with arbitrary width which is then
{formally} thermalized by convolution with a Maxwell-Boltzmann
velocity distribution function. One could argue, using Ford, Lewis,
and O'Connell's own logic~\cite{Ford_PLA01}, that the arbitrary
spatial extension and precise form of the initial Gaussian state on
the one hand, and its formal thermalization at an arbitrarily chosen
temperature~$\theta$ on the other, might not be independent and the
details need further justification. Indeed, the environmental
tempera\-ture typically determines the spatial extent of emerging
position-pointer states~\cite{Zurek_RMP03} and a generic mechanism,
such as photons randomly scattering off point particles with spatially
extended center-of-mass wave function, can lead to wave functions that
do not have Gaussian shape~\cite{Ole_PRA95}.

To avoid such ambiguities a possible physical implementation of the
scenario modelled here is introduced in
section~\ref{sec_steps}. Section~\ref{sec_Pure_States} illustrates
different cases of our scenario employing varied physical
parameters. In section~\ref{sec_thermal_states} we consider the impact
of temperature on the expected visibility of the interference pattern
and its loss (and resurgence) over time. Finally, in
section~\ref{sec_No_decoherence_Without}, we consider the `free case',
mostly discussed by Ford \emph{et~al.}, we will see that it can be
mapped onto the case of a mirror trapped harmonically. A harmonically
trapped thermal mirror shows periodic resurgence of the interference
pattern, we therefore conclude that ``decoherence without dissipation"
is absent even when adopting Ford and O'Connell's approach using
interference contrasts instead of non-diagonal density matrix
elements.

\IgnoreThis{\section{Steps of the scenario}\label{sec_steps}}

\IgnoreThis{\subsection{Trapped and cooled
    mirror}\label{subsec_trapped_cooled_mirror}}

A mirror with mass~$M$, initially trapped by a stiff harmonic
potential with spring constant~$K_0$ and cooled to
temperature~$\theta$~\cite{Corbitt_PRL07}, is described by the thermal
density matrix
\begin{eqnarray}
  \hat \rho_0 = \sum_{n=0}^{\infty}
  \frac{e^{-\frac{n\Theta_E}{\theta}}}{1-e^{-\frac{\Theta_E}{\theta}}}
  |\psi_n(x,0;K_0) \rangle\langle \psi_n(x,0;K_0)| \, .
  \label{eq_initial_mixed_state}
\end{eqnarray}
Here~$\Theta_E = \hbar \sqrt{K_0/M}/k_B$ is the associated
Einstein-temperature where $\hbar$ is Planck's and $k_B$~Boltzmann's
constant. In all numerical calculations we will set $\hbar=1$ and
$k_B=1$. For opto-mechanical oscillator masses ranging
from~$M=10^{-15}$ to~$10^{-10}$kg~\cite{Marquardt_Physics.2.40} and
recently reported large optical spring constants~$K_0 \approx
10^6$Nm${}^{-1}$~\cite{Corbitt_PRL07} the oscillator's resonance
frequency~$\Omega_0 = \sqrt{K_0/M}$ ranges from $3\cdot
10^{10}$s${}^{-1}$ to $10^{8}$s${}^{-1}$ and the associated
Einstein-temperature~$\Theta_E =\hbar \Omega_0 / k_B $ from 0.24 K to
0.76~mK. Such temperatures can be reached in optical cooling
setups~\cite{Corbitt_PRL07,Thompson_NAT08,Groeblacher_NATPhys09}. We
will see that the scheme presented here can tolerate temperatures up
to several~$\Theta_E$ of the trapping potential.

The normalized harmonic oscillator
eigen\-functions $\psi_n(x,0;K_0)= H ( n,{\frac {x}{\sigma_0}} )
{e^{-{\frac{{x}^{2}}{2{ \sigma_0}^{2}}}}} / \sqrt{ \sqrt{ \pi }
  \sigma_0 \,{2}^{n}n! } $ are para\-meter\-ized by the system's ground
state position standard deviation
\begin{eqnarray}
\sigma_0 = \frac{\sqrt{\hbar}}{\sqrt[4]{K_0 M}} . \label{eq_sigma}
\end{eqnarray}
For the values for $K_0$ and $M$ cited above, this position spread
is very small~($\sigma_0 \approx 10^{-14}$ to $5\cdot 10^{-16}$~m).
This is good to protect the oscillator from decoherence but makes it
hard to see interference fringes when momentum superposition states
are formed~\cite{Ole_quantumEntrained}.

\IgnoreThis{\subsection{Released mirror}\label{subsec_released_mirror}}

To be able to `imprint' the interference pattern associated with the
momentum superposition state~\cite{Ole_quantumEntrained} we will
therefore assume that the stiff trapping potential is suddenly
switched off so that the mirror is very weakly trapped or set free. In
the free case, the ensuing ballistic expansion of the stiff
potential's groundstate yields the `free groundstate' wave
function~\cite{Ole_FreeHOSC_arxiv_10}
\begin{eqnarray}
  \phi_0(x,t) & = & \frac {1 }{\sqrt{\sqrt{\pi} \sigma(t)}}
  \exp{\left[ {-\frac { x^2}{2 \sigma_0 \sigma(t)}} \right]}
  , \label{eq_phi_0_x_t} \\
  \mbox{where} \quad
  \sigma(t) & = & \sigma_0+i \frac{t\hbar}{\sigma_0 M} \label{eq_sigma_t}
\end{eqnarray}
is the time-dependent position spread which gives rise to the
{ballistic expansion velocity of the `released groundstate' (it is
larger for `released excited states')}
\begin{eqnarray}
v = \frac{\hbar}{\sigma_0 M} =
\frac{\sqrt{\hbar\sqrt{K_0}}}{{M}^{3/4}} . \label{eq_V_expansion}
\end{eqnarray}
For~$M=10^{-15}$ to~$10^{-10}$kg~\cite{Marquardt_Physics.2.40} and
optical spring constants~$K_0 \approx
10^6$Nm${}^{-1}$~\cite{Corbitt_PRL07} this velo\-city ranges from
$60~\mu$m$\cdot$s${}^{-1}$ to 10~nm$\cdot$s${}^{-1}$. In other words,
for a mirror mass of~$10^{-15}$kg, the free ground state \wf expands
to an optical wavelength size of roughly 500 nm within an expansion
time of approximately $9$~ms.

\IgnoreThis{\subsection{Kicked thermal
mirror}\label{subsec_kicked_thermal_mirror}}

In order to achieve the momentum superposition, it is suggested to use
an interferometric setup featuring a half-silvered beam splitter that
splits a single photon (generated in spontaneous parametric
down-conversion~\cite{HongOuMandel87}) into two equally strong partial
waves. These partial waves are directed onto either side of the mirror
and, upon reflection, transfer their momentum recoil to the mirror in a
desirable fashion: namely, after tracing over the photon wave function
the mirror's wave function becomes multiplied with an effective
kick-factor~\cite{Ole_quantumEntrained}
\begin{equation}
{\cal K} = -i \sin(2 \kappa x - \frac{\phi}{2}) \; ,
\label{eq_Kick_factor}
\end{equation}
where the relative phase~$\phi$ is fixed by the interferometric setup
and the port the photon takes~\cite{Ole_quantumEntrained}. Here, it is
assumed that the mirror has unit reflectivity and $\kappa$ is the
photon's effective wavenumber, associated with its normal momentum
component, for more details and the case of imperfect mirrors
see~\cite{Ole_quantumEntrained}.

The `released eigenstates', spawned by the eigenfunctions~$\psi_n$
of the stiff potential, are labelled~$\Psi_n(x,t;0,\tau,k,K_0)$;
here~$\tau<0$ denotes the time when the mirror is released from the
stiff potential $K_0 x^2/2$ into a weak potential~$k x^2/2$ (or set
free:~$k=0$, in which case~$\Psi_0(x,0;0,-t,0,K_0)=\phi_0(x,t)$ of
Eq.~(\ref{eq_phi_0_x_t})).

Then, at time~$t=0$, the photon kicks the mirror into a
momentum-superposition state. Through the momentum kick the wave
functions 
become combined to form the desired momentum superposition
states~$\Upsilon_n \propto {\cal K} \Psi_n$
\begin{eqnarray}
\Upsilon_n(x,t;p_\gamma,\tau,\phi,k,K_0)& = & {\cal N}_n
[\Psi_n(x,t;p_\gamma,\tau,k,K_0)
\nonumber \\
&- & e^{i\phi}\Psi_n(x,t;-p_\gamma,\tau,k,K_0)] \, . \quad
\label{eq_final_pure_state}
\end{eqnarray}
Here, ${\cal N}_n$ is a normalization constant that absorbs overall
phase factors and primarily depends on the size of the photons'
momentum transfers~$p_\gamma=2 \hbar \kappa$,
compare~Eq.~(\ref{eq_Kick_factor}), and the position spread of each
state~$\Psi_n$ at the time (t=0) when the photon-kick occurs. The
thermal momentum superposition state thus has the form
\begin{eqnarray}
  \hat \rho(x,t;p_\gamma,\tau,\phi,k,K_0) = \sum_{n=0}^{N}
  \frac{e^{-\frac{n\Theta_E}{\theta}}}{1-e^{-\frac{\Theta_E}{\theta}}}
  |\Upsilon_n  \rangle\langle \Upsilon_n| \, .
\label{eq_final_mixed_state}
\end{eqnarray}
The states~$\Upsilon_n$ up to 13-th order have been determined in this
work, so, the cut-off~$N=13$. Upon inclusion of the `free' kicked
groundstate~$\Upsilon_0$, these lowest fourteen oscillator states
allow us to model the stiff oscillator at temperatures of up to three
times the Einstein-temperature: $\sum_{n=14}^{\infty}e^{n/3.0}<0.01$,
since this accounts for 99\% of the probability distribution for
temperature~$\theta =3.0 \cdot \Theta_E$ at which a harmonic
oscillator has roughly 99.08\% of the classical heat
capacity~$\frac{2}{2} k_B$. At such temperatures, the character of the
system can be considered to be between quantum and classical.

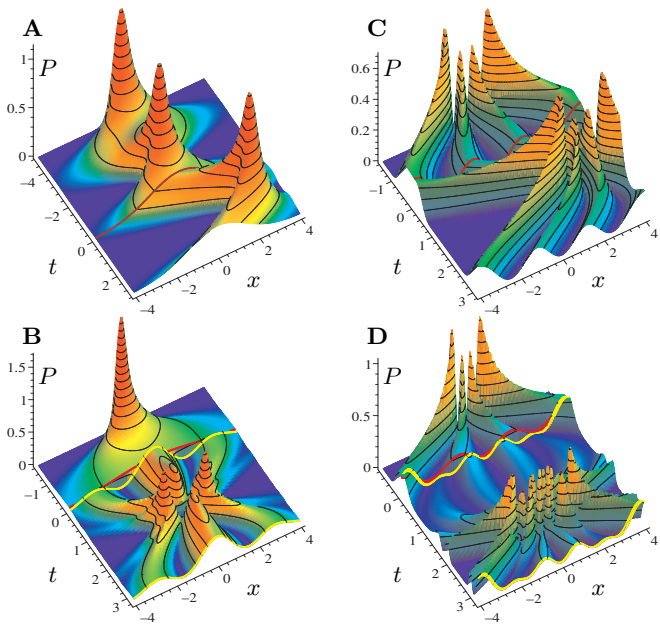
\begin{figure}[t]
\begin{center}
  \begin{minipage}[b]{0.47\linewidth}
     \includegraphics[width=1\linewidth,height=1\linewidth]{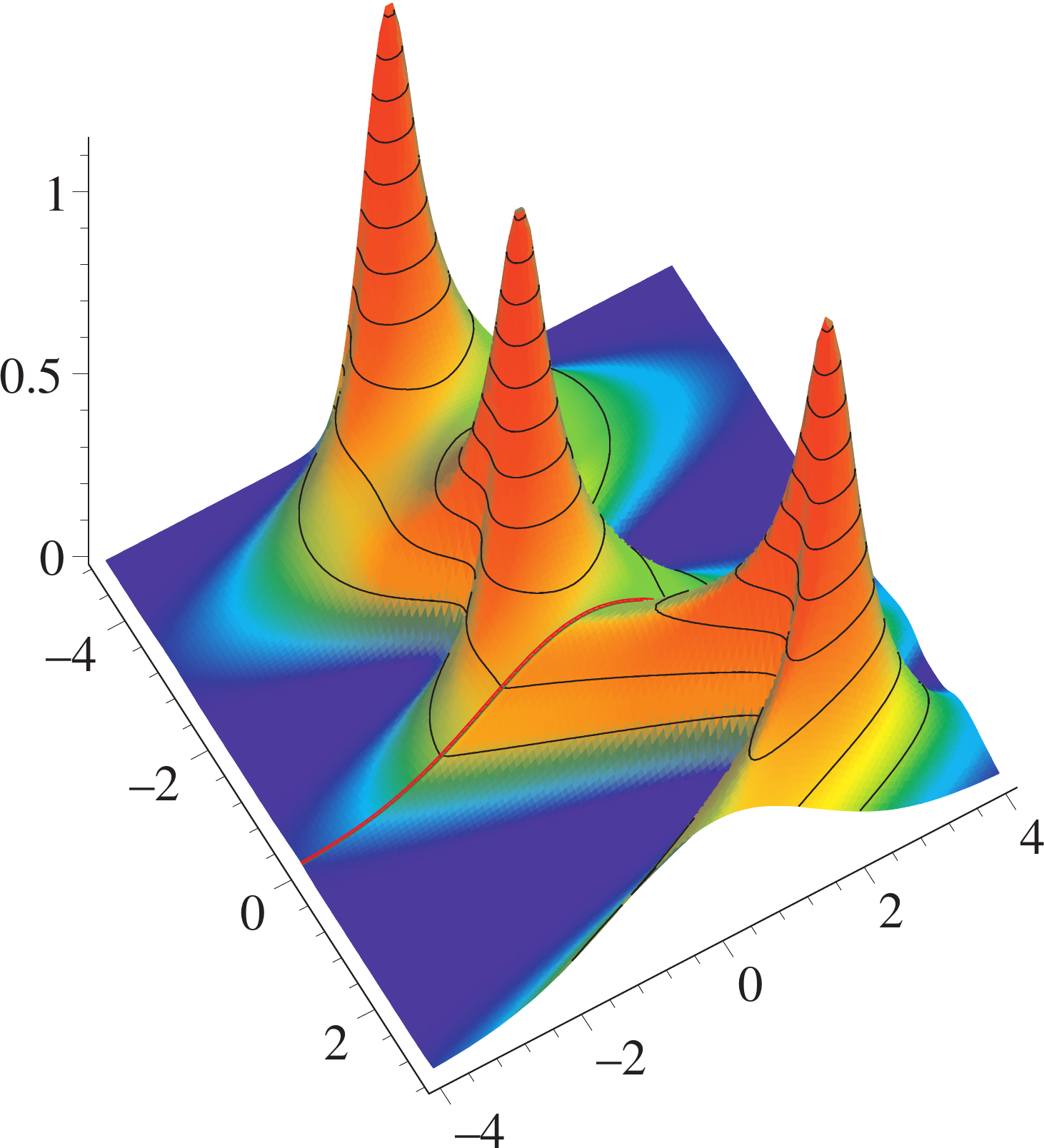}
     \put(-110,105){\rotatebox{0}{\mbox{\bf A}}}
     \put(-104,90){\rotatebox{0}{$P$}}
     \put(-100,15){\rotatebox{0}{$t$}}
     \put(-25,10){\rotatebox{0}{\mbox{$x$}}}
  \end{minipage}
  \hspace{0.035\linewidth}
  \begin{minipage}[b]{0.47\linewidth}
     \includegraphics[width=1\linewidth,height=1\linewidth]{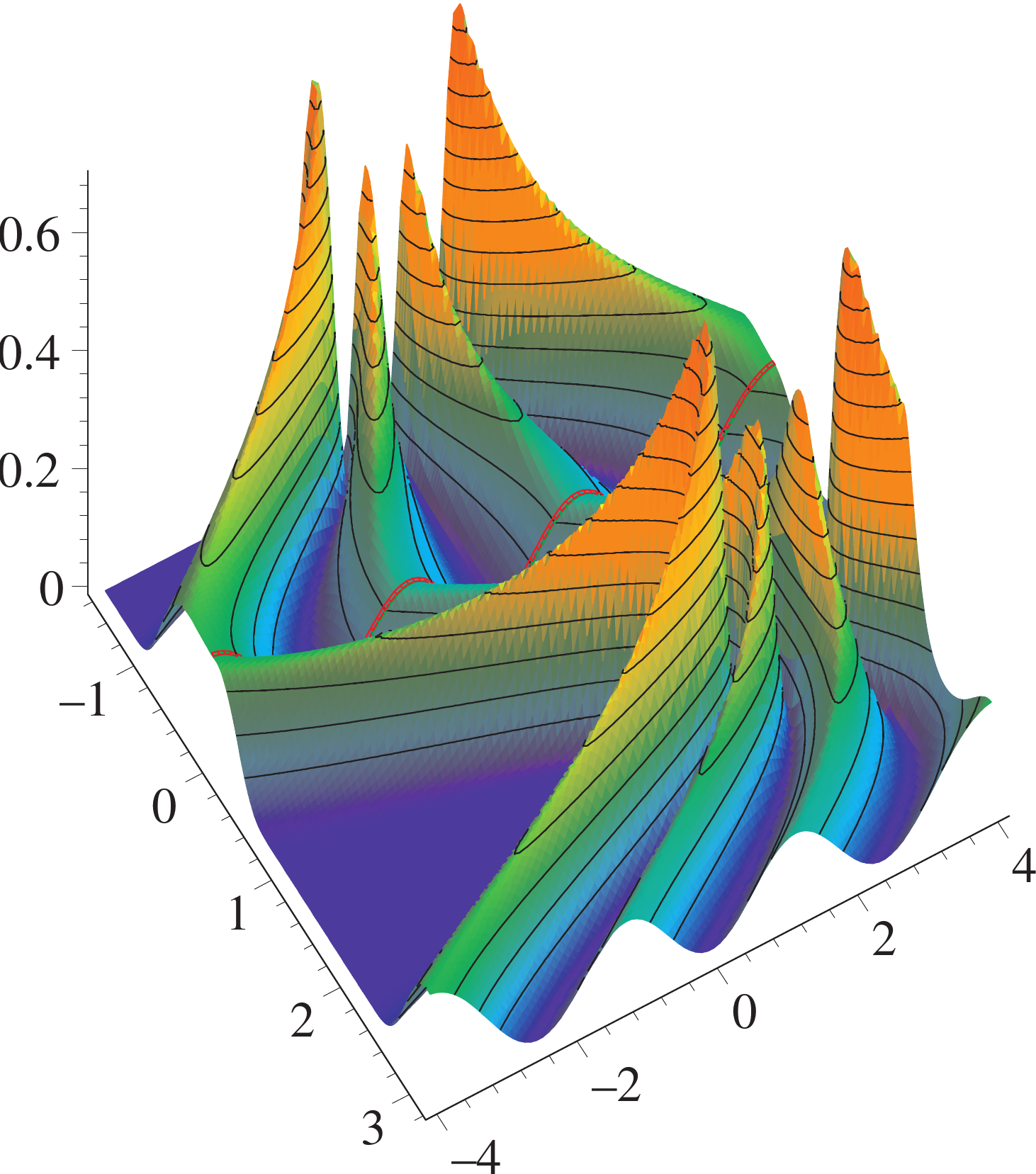}
     \put(-110,105){\rotatebox{0}{\mbox{\bf C}}}
     \put(-104,90){\rotatebox{0}{$P$}}
     \put(-100,15){\rotatebox{0}{$t$}}
     \put(-25,10){\rotatebox{0}{\mbox{$x$}}}
  \end{minipage}
\\
  \begin{minipage}[b]{0.47\linewidth}
     \includegraphics[width=1\linewidth,height=1\linewidth]{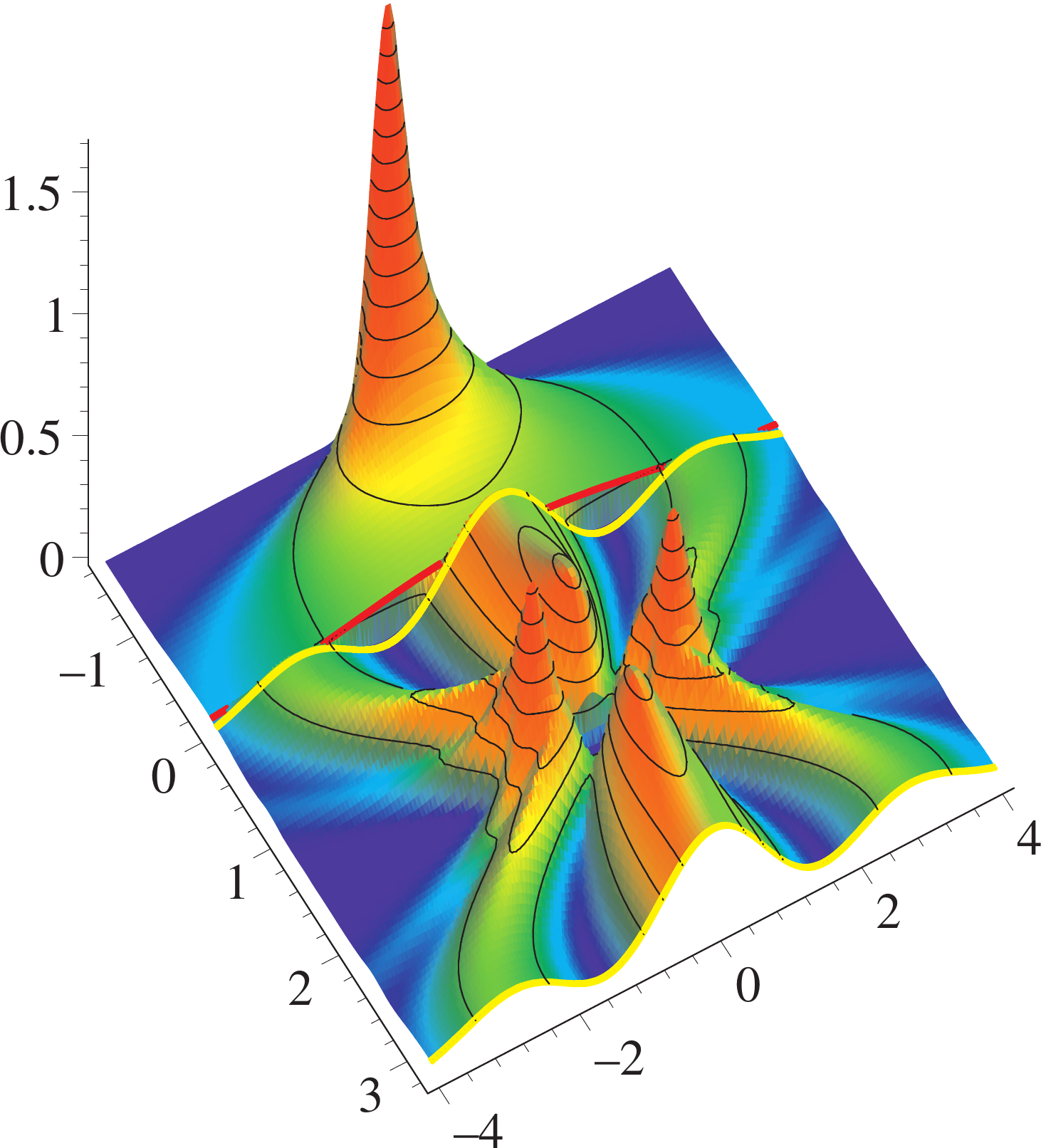}
     \put(-110,105){\rotatebox{0}{\mbox{\bf B}}}
     \put(-104,90){\rotatebox{0}{$P$}}
     \put(-100,15){\rotatebox{0}{$t$}}
     \put(-25,10){\rotatebox{0}{\mbox{$x$}}}
  \end{minipage}
  \hspace{0.035\linewidth}
  \begin{minipage}[b]{0.47\linewidth}
     \includegraphics[width=1\linewidth,height=1\linewidth]{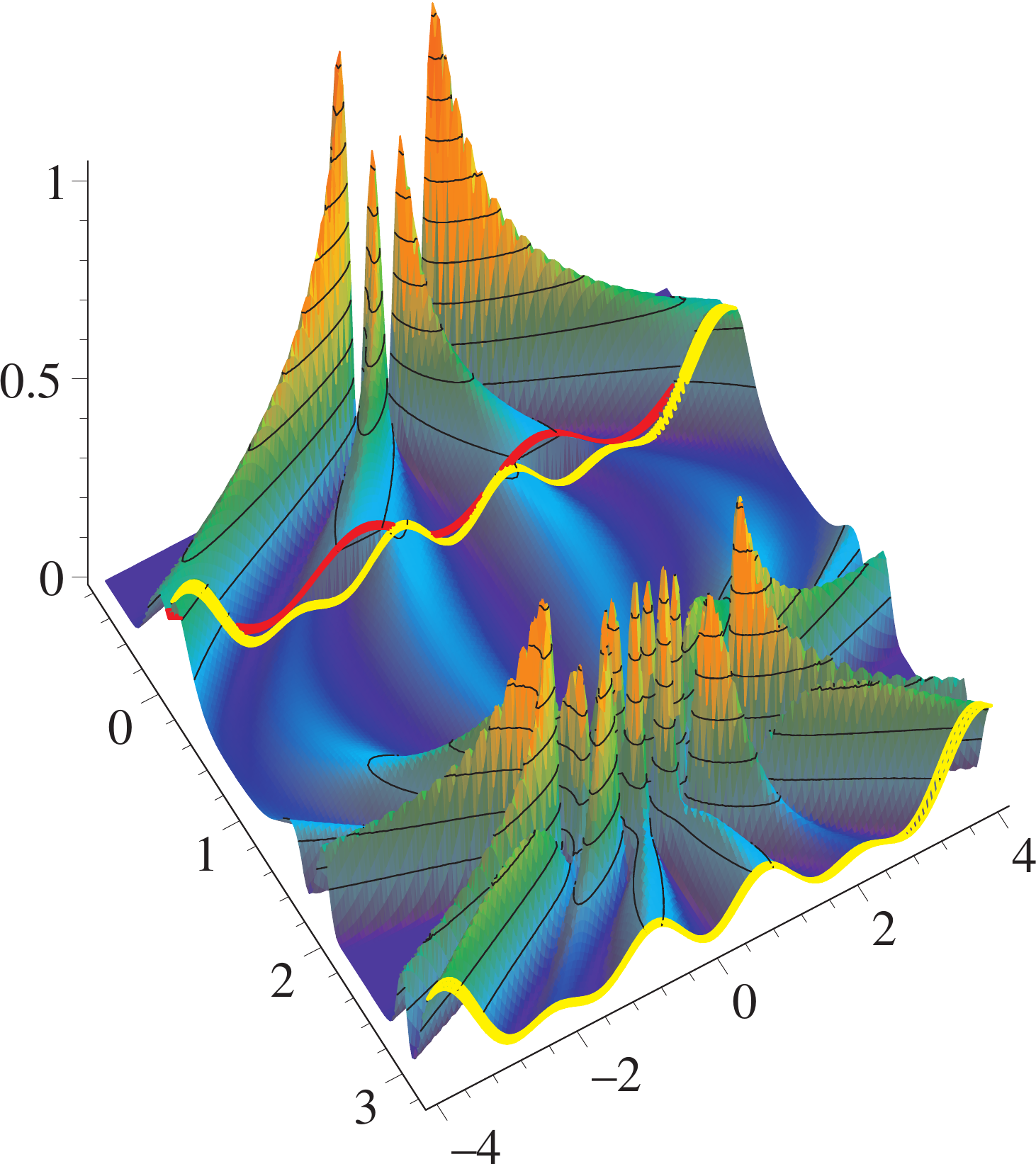}
     \put(-110,105){\rotatebox{0}{\mbox{\bf D}}}
     \put(-104,90){\rotatebox{0}{$P$}}
     \put(-100,15){\rotatebox{0}{$t$}}
     \put(-25,10){\rotatebox{0}{\mbox{$x$}}}
  \end{minipage}
\end{center}
\caption{Probability densities $P(x,t)$ of quantum harmonic oscillator
  with mass~$M=1$ released into a weak harmonic potential
  ($\tau=-T/4$,~$k=1$) and subsequently momentum-kicked at time zero
  ($\phi=\pi$, for how fixed phases arise in entrained measurements
  see~\cite{Ole_quantumEntrained}).  Plots {\bf A} and {\bf B} show
  released and kicked ($p_\gamma = 1$) groundstate~$\Psi_0$ with
  different initial conditions ($K_0=2^4$ and single kick, versus,
  $K_0=3^4$ and momentum superposition kick). Similarly, plots {\bf C}
  and {\bf D} show released and kicked~($p_\gamma = 2.5$) third
  excited state~$\Psi_3$ with $K_0=2^4$ and~$\tau=-T/4$, and $K_0=3^4$
  and~$\tau=-T/8$, respectively.} \label{fig_pure_states}
\end{figure}

\section{Dynamics of pure states}\label{sec_Pure_States}

We now study the dynamics of the groundstate, which is an even
function in position, and the third excited state, which is odd.

Figure~\ref{fig_pure_states}~{\bf A} shows the groundstate released
from the stiff oscillator potential at $\tau = -3 T/4$ where $T=2\pi
\sqrt{M/k}=2\pi/\omega$ is the mirror's period time. It plays out as
an underlying squeezed and anti-squeezed breathing motion in
evidence in all plots of Figure~\ref{fig_pure_states}. At time zero
the mirror is kicked to the right and so shows a combination of
breathing and oscillation. Plot~{\bf B} shows the same scenario
except for the facts that the release from the stiff potential
happens at time $\tau = - T/4$ and the momentum-kick is in
superposition. The formation of the superposition is highlighted by
the red line immediately before the momentum kick and the yellow
line when it happens. The same yellow profile line has been moved
downstream and shows that intensity patterns re-form every half
period.

Plots~\ref{fig_pure_states}~{\bf C} and {\bf D} show analogous
scenarios for the third excited state released from the stiff
potential at times $\tau = - T/4$ and $-T/8$ respectively. Plots~{\bf
  B} and~{\bf D} illus\-trate that the interference pattern of the
momentum superposition is imprinted onto the underlying spatial
probability distribution of the state with full contrast.

\begin{figure}[t]
\begin{center}
  \begin{minipage}[b]{0.47\linewidth}
     \includegraphics[width=1\linewidth,height=1\linewidth]{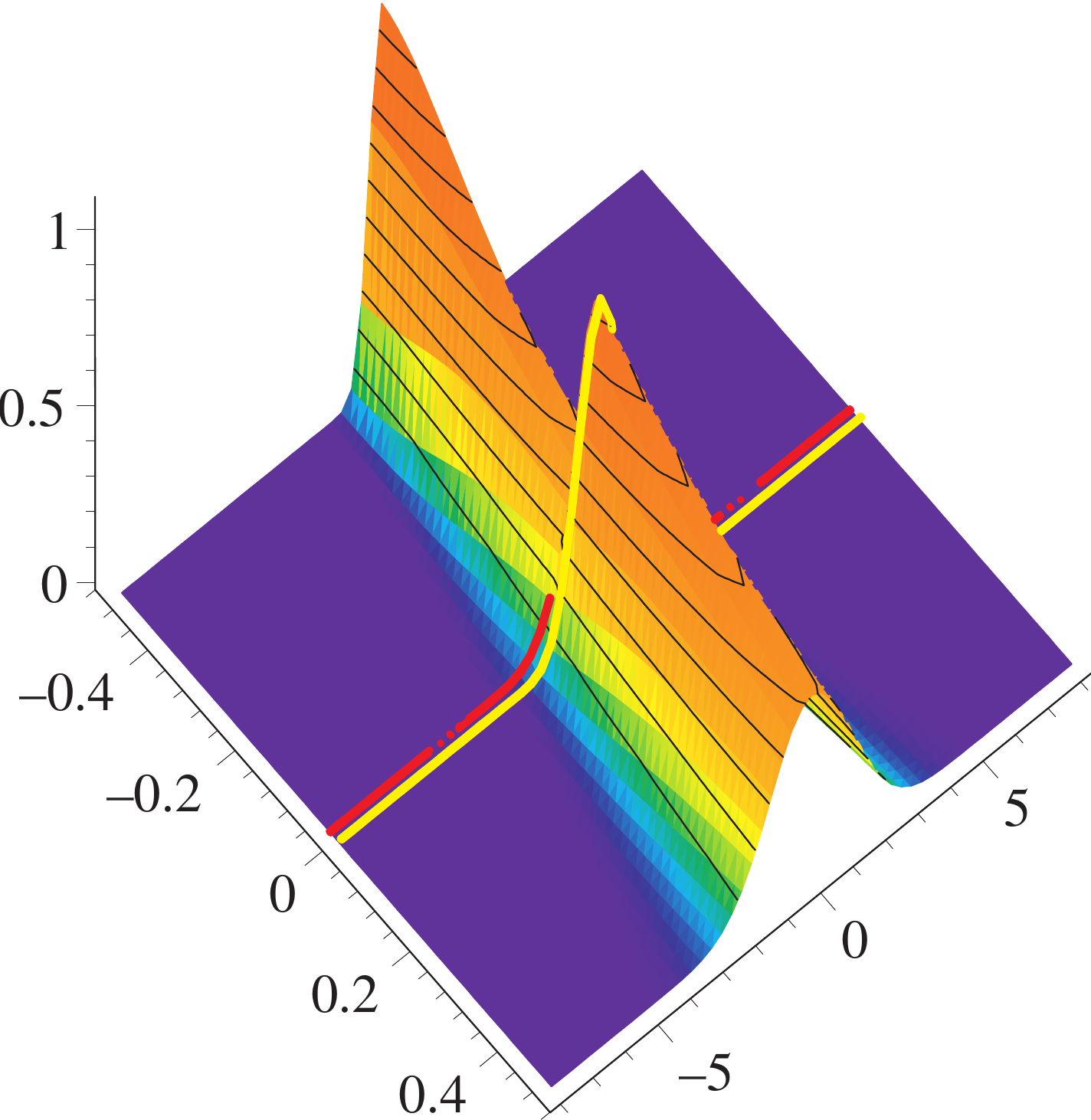}
     \put(-110,105){\rotatebox{0}{\mbox{\bf A}}}
     \put(-104,90){\rotatebox{0}{$P$}}
     \put(-95,15){\rotatebox{0}{$t$}}
     \put(-10,15){\rotatebox{0}{\mbox{$x$}}}
  \end{minipage}
  \hspace{0.035\linewidth}
  \begin{minipage}[b]{0.47\linewidth}
     \includegraphics[width=1\linewidth,height=1\linewidth]{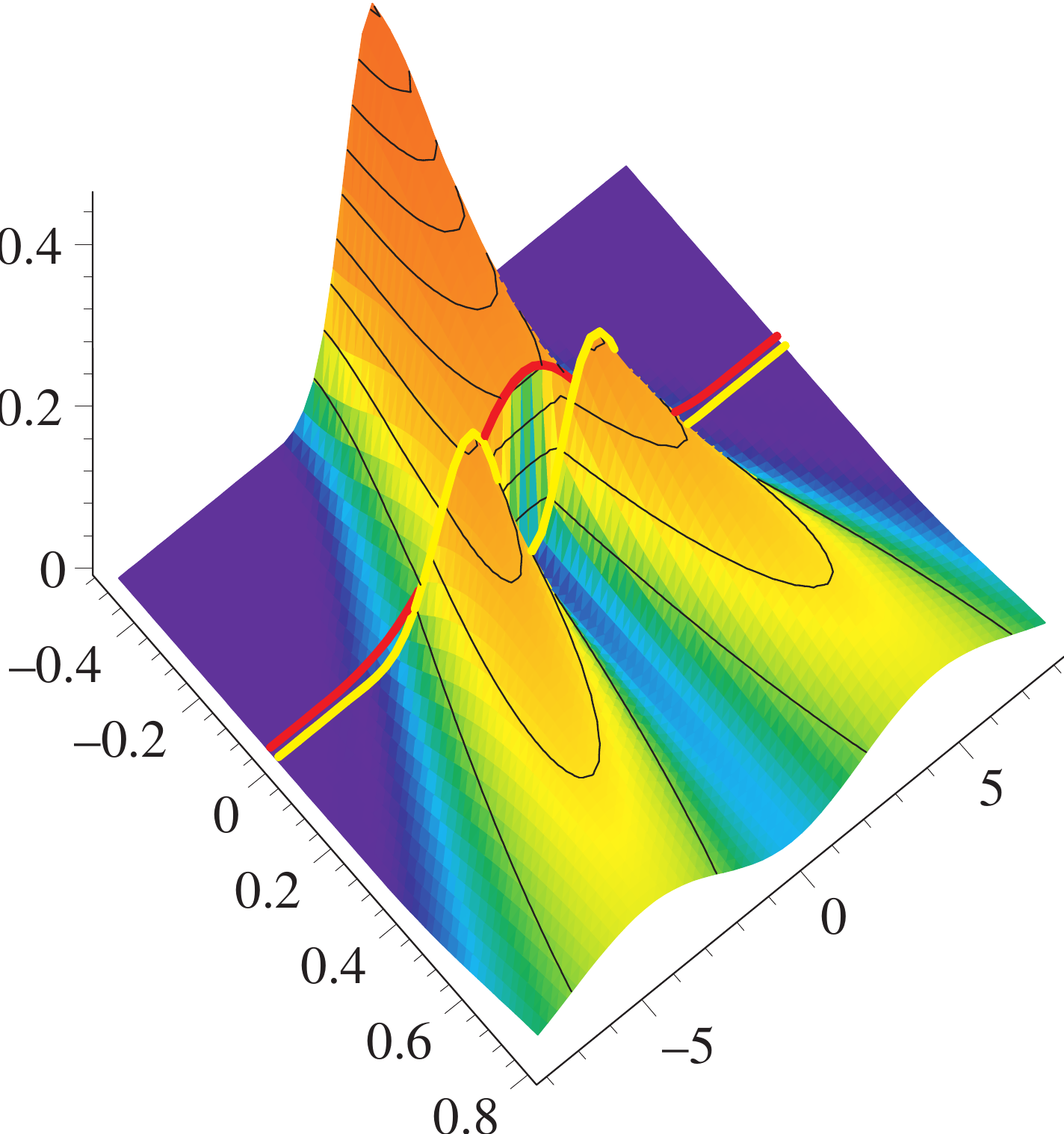}
     \put(-110,105){\rotatebox{0}{\mbox{\bf C}}}
     \put(-104,90){\rotatebox{0}{$P$}}
     \put(-95,15){\rotatebox{0}{$t$}}
     \put(-10,15){\rotatebox{0}{\mbox{$x$}}}
  \end{minipage}
\\
  \begin{minipage}[b]{0.47\linewidth}
     \includegraphics[width=1\linewidth,height=1\linewidth]{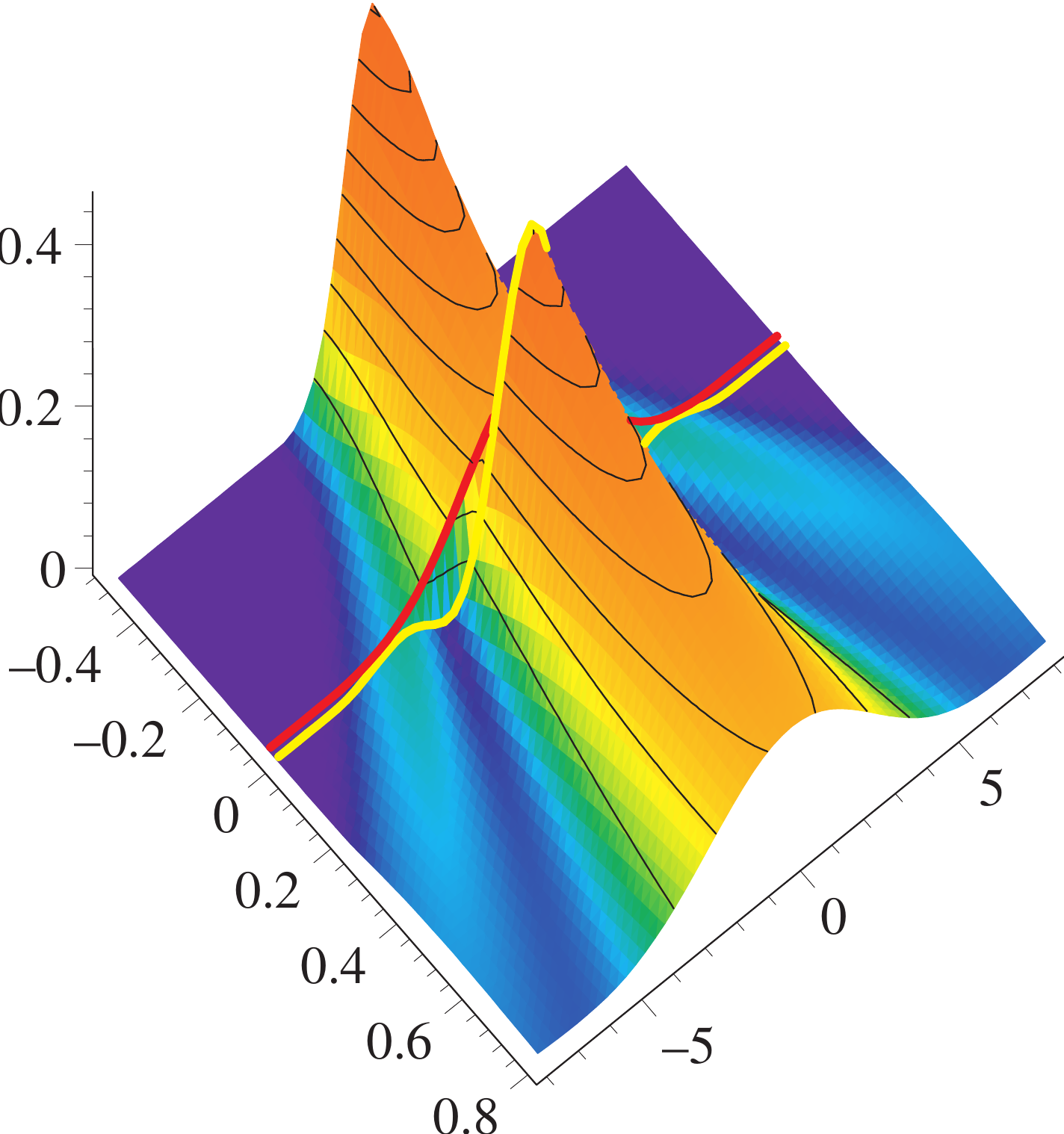}
     \put(-110,105){\rotatebox{0}{\mbox{\bf B}}}
     \put(-104,90){\rotatebox{0}{$P$}}
     \put(-95,15){\rotatebox{0}{$t$}}
     \put(-10,15){\rotatebox{0}{\mbox{$x$}}}
  \end{minipage}
  \hspace{0.035\linewidth}
  \begin{minipage}[b]{0.47\linewidth}
     \includegraphics[width=1\linewidth,height=0.8\linewidth]{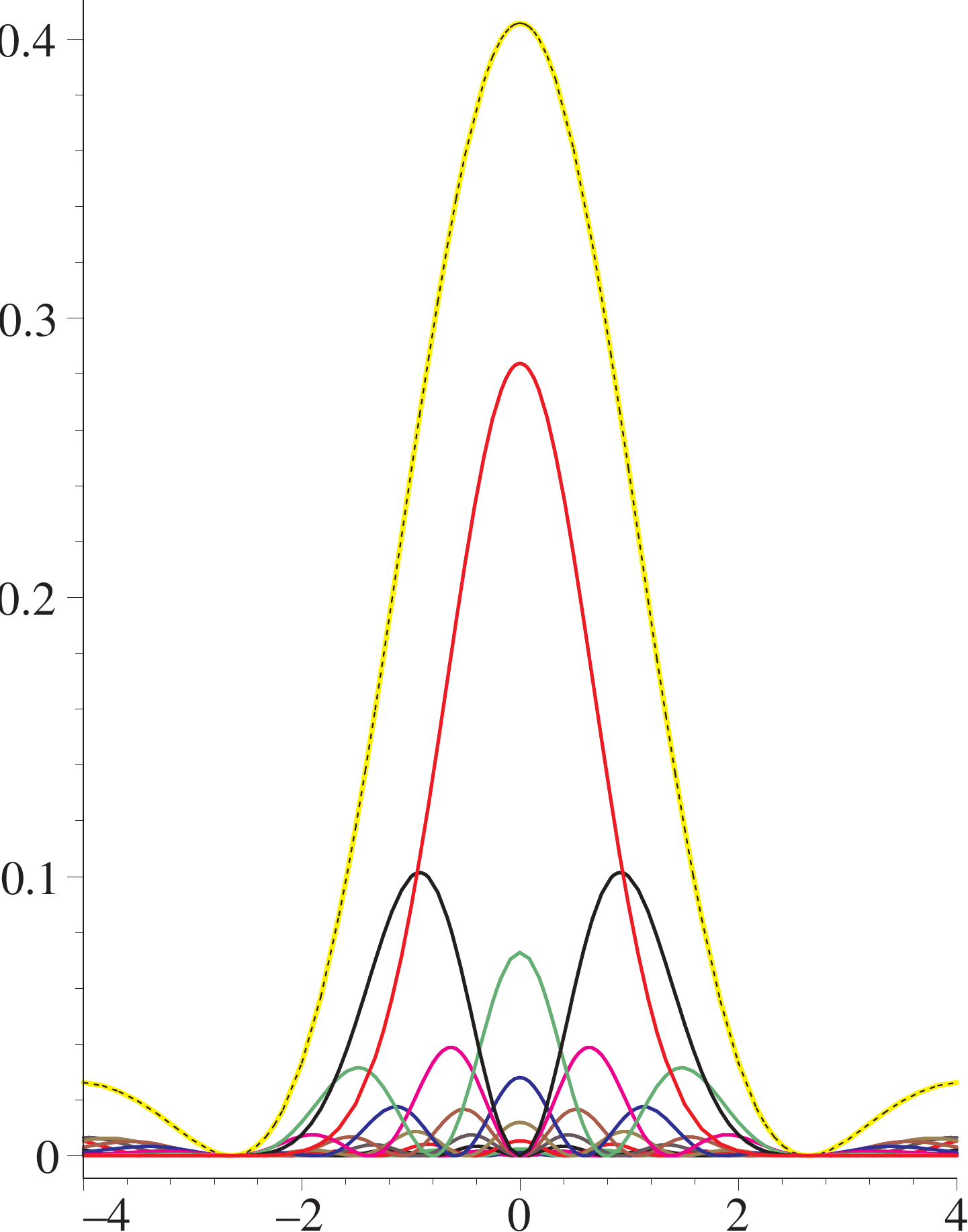}
     \put(-102,89){\rotatebox{0}{$P$}}
     \put(-10,-5){\rotatebox{0}{\mbox{$x$}}}
     \put(-110,105){\rotatebox{0}{\mbox{\bf D}}}
  \end{minipage}
\end{center}
\caption{Probability densities $P(x,t) =
  \rho(x,t;\frac{3}{5},-\frac{1}{2},0,K_0)$ of quantum harmonic
  oscillator with mass~$M=1$ set free~($k=0$) at time $\tau = -1/2$
  from stiff oscillator with spring constant $K_0=2^4$ and
  momentum-superposition--kicked at time zero: $p_\gamma = 3/5$. Plot
  {\bf A} shows mirror cooled almost to ground state $\theta =
  \Theta_E/3$.  It demonstrates that the small extension of the wave
  packet is not wide enough to imprint a full interference
  pattern~($\phi=\pi$), although a slimming and heightening of the
  probability distribution can be seen.  Plot {\bf B} same as {\bf A}
  for $\theta = 3 \Theta_E$.  Clearly the wave packet is wider and
  expands faster at this higher temperature allowing for a full
  imprint of the inter\-ference pattern. Plots {\bf C}, same as {\bf
    B} with centered inter\-ference mini\-mum~($\phi = 0$). The
  dephasing of the inter\-ference pattern in this case is greater than
  in~{\bf B} because fewer states contribute to dephasing on the
  fringe than at the center~$x=0$.  Plot~{\bf D}: Under the dotted
  envelope~$P(x,0)$ the thermally weighted contributions of the
  `released eigenstates' are shown, demonstrating that fewer
  eigenstates contribute to interference on the fringe (case {\bf B})
  than at the center (case~{\bf C}). } \label{fig_thermal_states_free}
\end{figure}
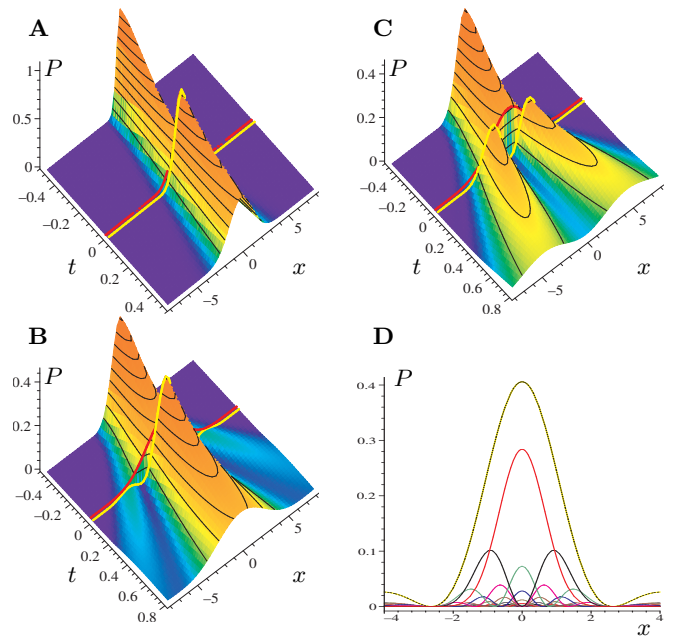

\section{Dynamics of thermal states}\label{sec_thermal_states}

At the temperatures considered here, ranging from zero up to three
times the Einstein temperature, the interference with full contrast
can be observed for all settings of the system's parameters such as
spring constants, mass, and photon momenta, provided that the
released mirror wave functions had time to ballistically expand
sufficiently widely to accommodate an imprint at the effective
wavelength~\cite{Ole_quantumEntrained} of the photon recoil when it
is being kicked into the superposition state,
compare~Fig.~\ref{fig_thermal_states_free}~{\bf A} versus~{\bf B}.
In other words, wave packet expansion and the momentum-superposition
interference imprinting operations do not commute. Each wave
function~$\Upsilon_n$ is by itself fully coherent and the momentum
superposition imprints occur at the same spatial positions for all
of them. So, the effects add up to an interference pattern with full
contrast. This implies that incoherent mixtures consisting of states
spread out wide enough to carry the momentum superposition imprint
can be endowed with an interference pattern when using the
entrainment procedure of reference~\cite{Ole_quantumEntrained}. This
observation, which also applies to states with temperatures much
higher than studied here, is one of the central messages of this
work.

Note that the dephasing of the imprinted interference pattern does
not only depend on the temperature but also its location. Further
out from the center fewer `released eigenstates'~$\Upsilon_n$
contribute to the thermal wave packet and therefore dephasing is
slowed down there as compared to dephasing at the center, $x=0$,
compare~Fig.~\ref{fig_thermal_states_free}~{\bf B} with~{\bf C}. On
the fringes the interference gets washed out more slowly.

Additionally, more highly excited `released eigenstates' have a wider
spread and higher ballistic expansion velocity, they therefore can
carry an interference imprint, when lower lying `released eigenstates'
might not be able to do this, see
Fig.~\ref{fig_thermal_states_free}~{\bf B} as opposed to~{\bf A}.
Surprisingly, higher temperatures can help in generating an
interference pattern imprint because they expand the width of the
mirror's wavepacket.

Since they have different energies the various wave
functions~$\Upsilon_n$ evolve at different rates and therefore
mutually dephase thus washing out the interference imprint. The
general scenario is varied and its detailed quantification is beyond
the level of this work. We would, however, like to point out that
the breathing motion underlying the dynamics in the trapped case
(see breathing motion of pure states in Fig.~\ref{fig_pure_states})
is least pronounced at every turning point or odd quarter period
point: $(2n+1)T/4=(2n+1) 2 \pi / (4 \omega)$ counted from the moment
of release~$\tau$. This is confirmed by
Fig.~\ref{fig_thermal_states_trapped}~{\bf B} and~{\bf C} which show
slowed dephasing since the momentum-superposition kick happens at
the quarter time ($\tau = -T/4$, $t=0$).

\subsection{Benchmark for dephasing function}\label{subsec_dephasing}

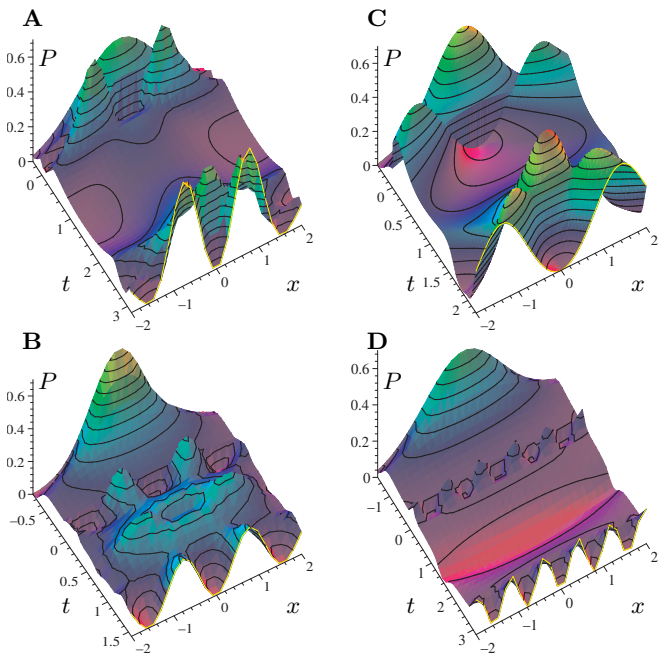
\begin{figure}[t]
\begin{center}
  \begin{minipage}[b]{0.47\linewidth}
     \includegraphics[width=1\linewidth,height=1\linewidth]{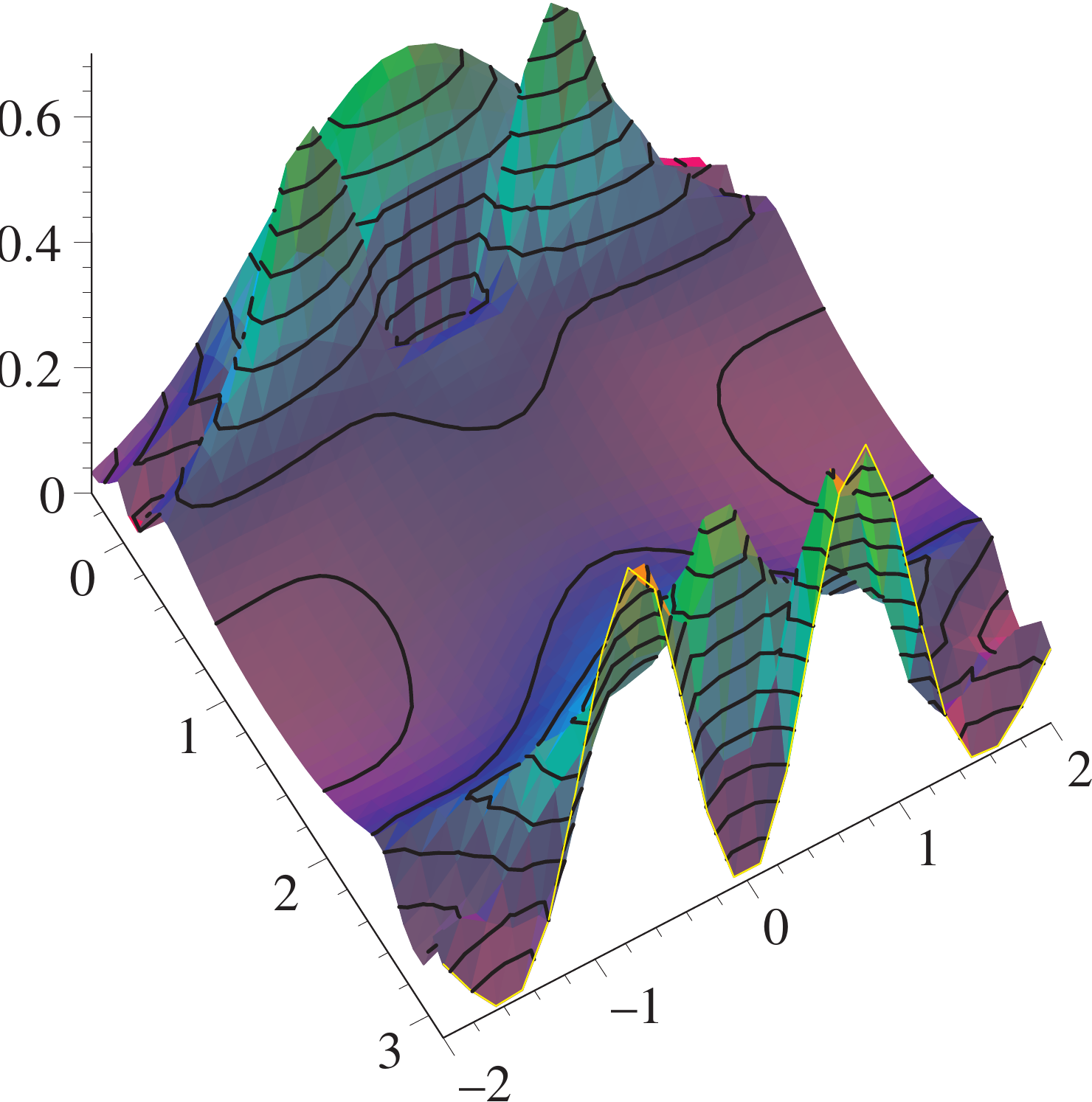}
     \put(-110,115){\rotatebox{0}{\mbox{\bf A}}}
     \put(-104,100){\rotatebox{0}{$P$}}
     \put(-95,15){\rotatebox{0}{$t$}}
     \put(-10,15){\rotatebox{0}{\mbox{$x$}}}
  \end{minipage}
  \hspace{0.035\linewidth}
  \begin{minipage}[b]{0.47\linewidth}
     \includegraphics[width=1\linewidth,height=1\linewidth]{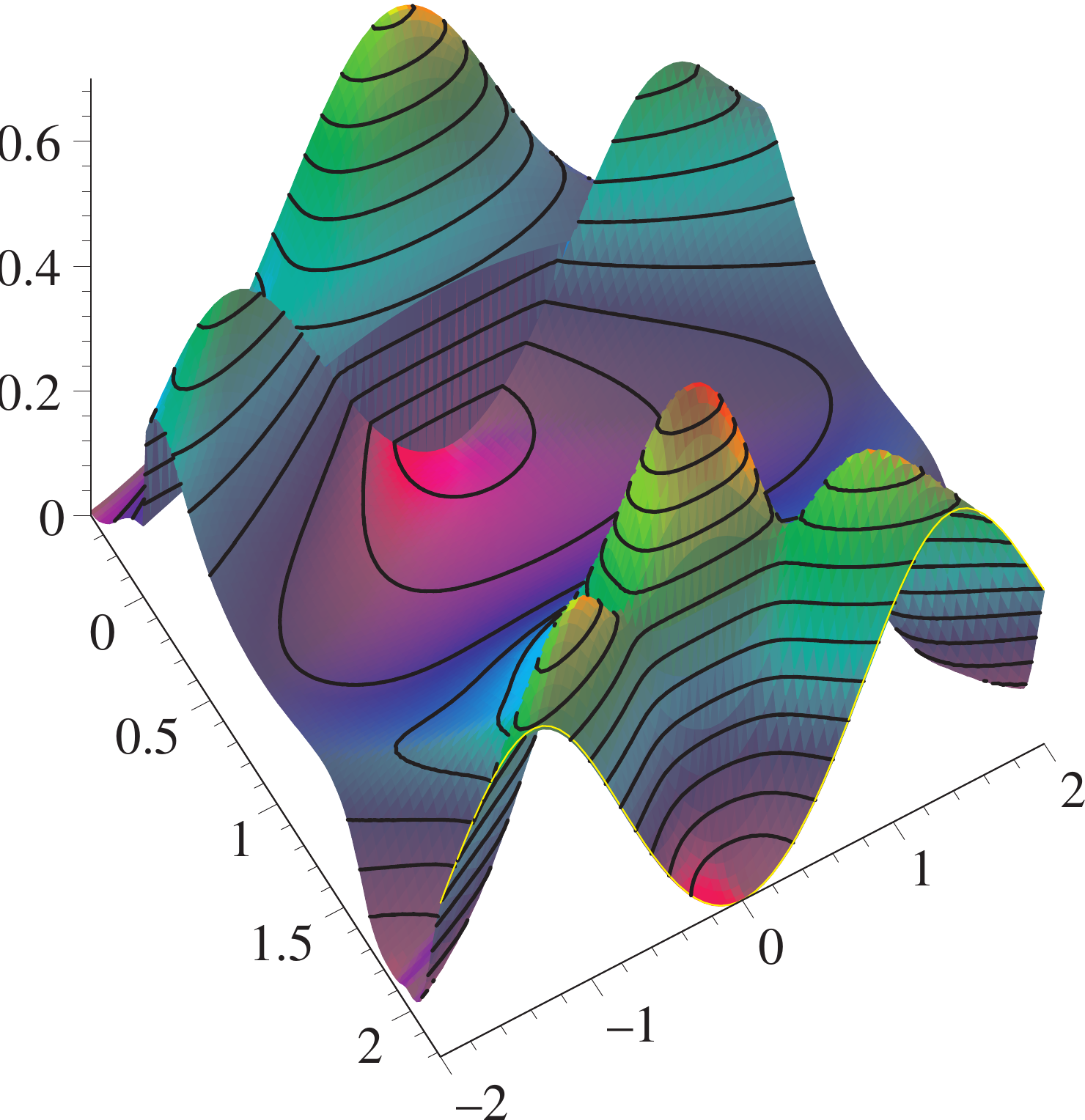}
     \put(-110,115){\rotatebox{0}{\mbox{\bf C}}}
     \put(-104,100){\rotatebox{0}{$P$}}
     \put(-95,15){\rotatebox{0}{$t$}}
     \put(-10,15){\rotatebox{0}{\mbox{$x$}}}
  \end{minipage}
\\
  \begin{minipage}[b]{0.47\linewidth}
     \includegraphics[width=1\linewidth,height=1\linewidth]{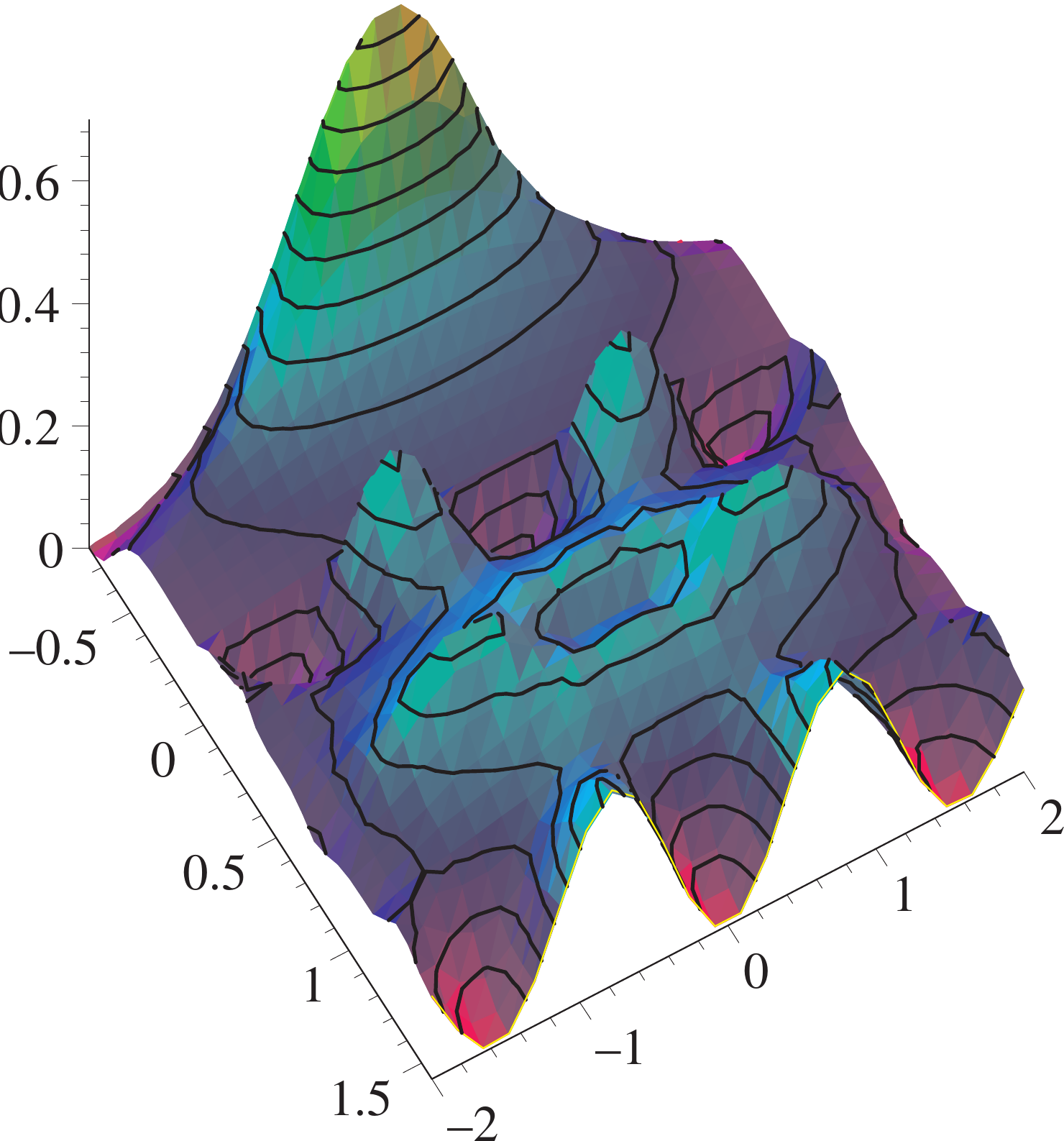}
     \put(-110,115){\rotatebox{0}{\mbox{\bf B}}}
     \put(-104,100){\rotatebox{0}{$P$}}
     \put(-95,15){\rotatebox{0}{$t$}}
     \put(-10,15){\rotatebox{0}{\mbox{$x$}}}
  \end{minipage}
  \hspace{0.035\linewidth}
  \begin{minipage}[b]{0.47\linewidth}
     \includegraphics[width=1\linewidth,height=1\linewidth]{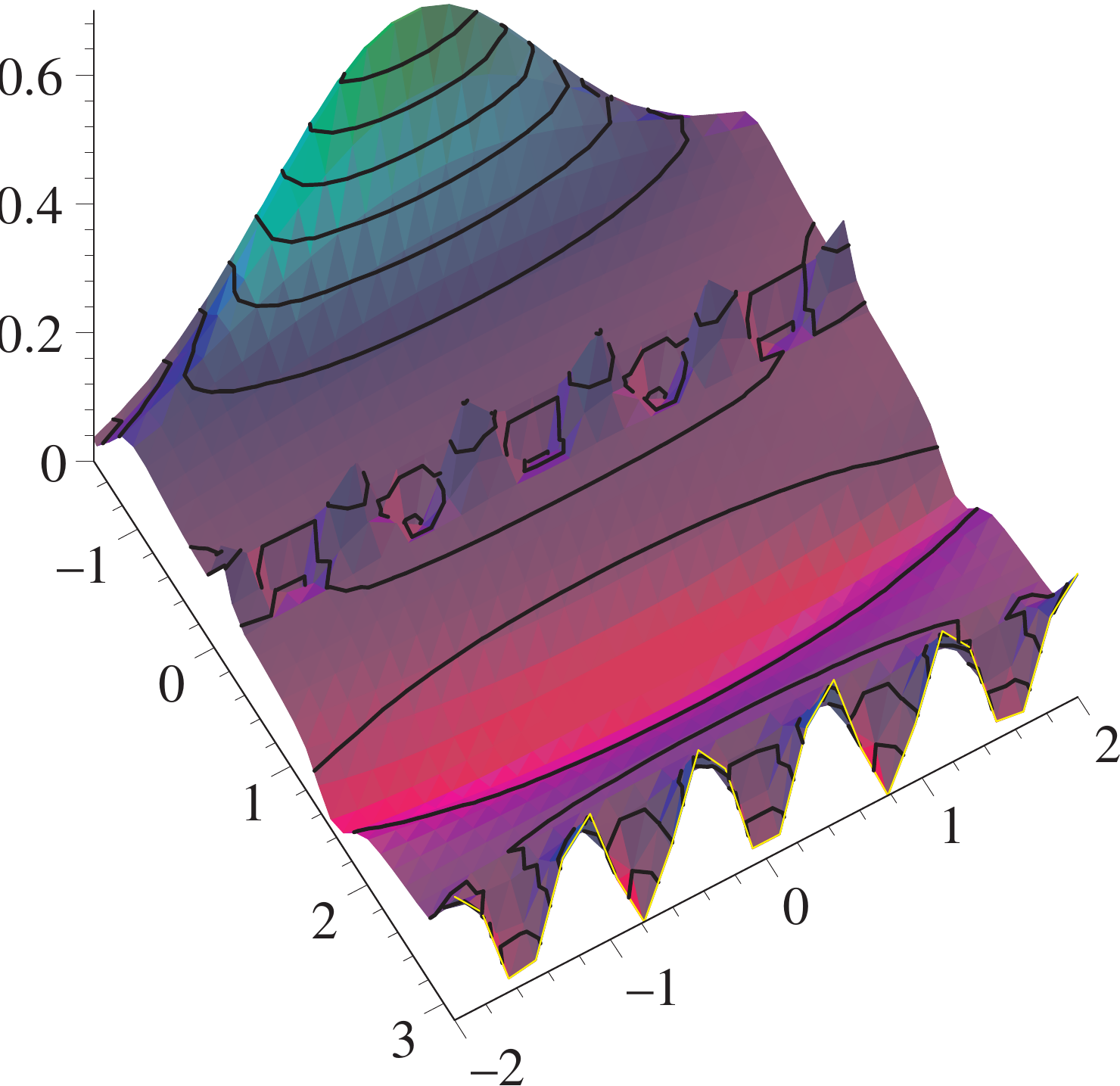}
     \put(-110,115){\rotatebox{0}{\mbox{\bf D}}}
     \put(-104,100){\rotatebox{0}{$P$}}
     \put(-95,15){\rotatebox{0}{$t$}}
     \put(-10,15){\rotatebox{0}{\mbox{$x$}}}
  \end{minipage}
\end{center}
\caption{Probability densities $P(x,t)$ of quantum harmonic oscillator
  with mass~$M=1$ released from stiff oscillator potential into weak
  potential and kicked into momentum superposition state at time
  $t=0$: Plot {\bf A} $\theta=3\Theta_E$, $K=16$, $k=1$, $p_\gamma=2$ and
  $\tau=-T/20$; {\bf B} $\theta=3\Theta_E$, $K=64$, $k=4$, $p_\gamma=2$ and
  $\tau=-T/4$; {\bf C} $\theta=2.815 \Theta_E$, $K=32$, $k=2$, $p_\gamma=1$ and
  $\tau=-T/10$; {\bf D} $\theta=3 \Theta_E$, $K=16$, $k=1$, $p_\gamma=4$ and
  $\tau=-T/4$. It can be seen that the interference patterns always re-form
  with full contrast after half a period. Moreover, the time over which the
  interference pattern persisters is lengthened when one kicks the mirror
  at the point of greatest expansion rather than
  earlier (see~{\bf B} as opposed to~{\bf A}). The dephasing of the
  interference pattern is weaker for longer wavelength imprints
  (smaller momenta), compare~{\bf A} and~{\bf B}, with~{\bf C} and~{\bf
  D}.
} \label{fig_thermal_states_trapped}
\end{figure}

\begin{figure}[t]
\begin{center}
  \begin{minipage}[b]{0.47\linewidth}
     \includegraphics[width=1\linewidth,height=1\linewidth]{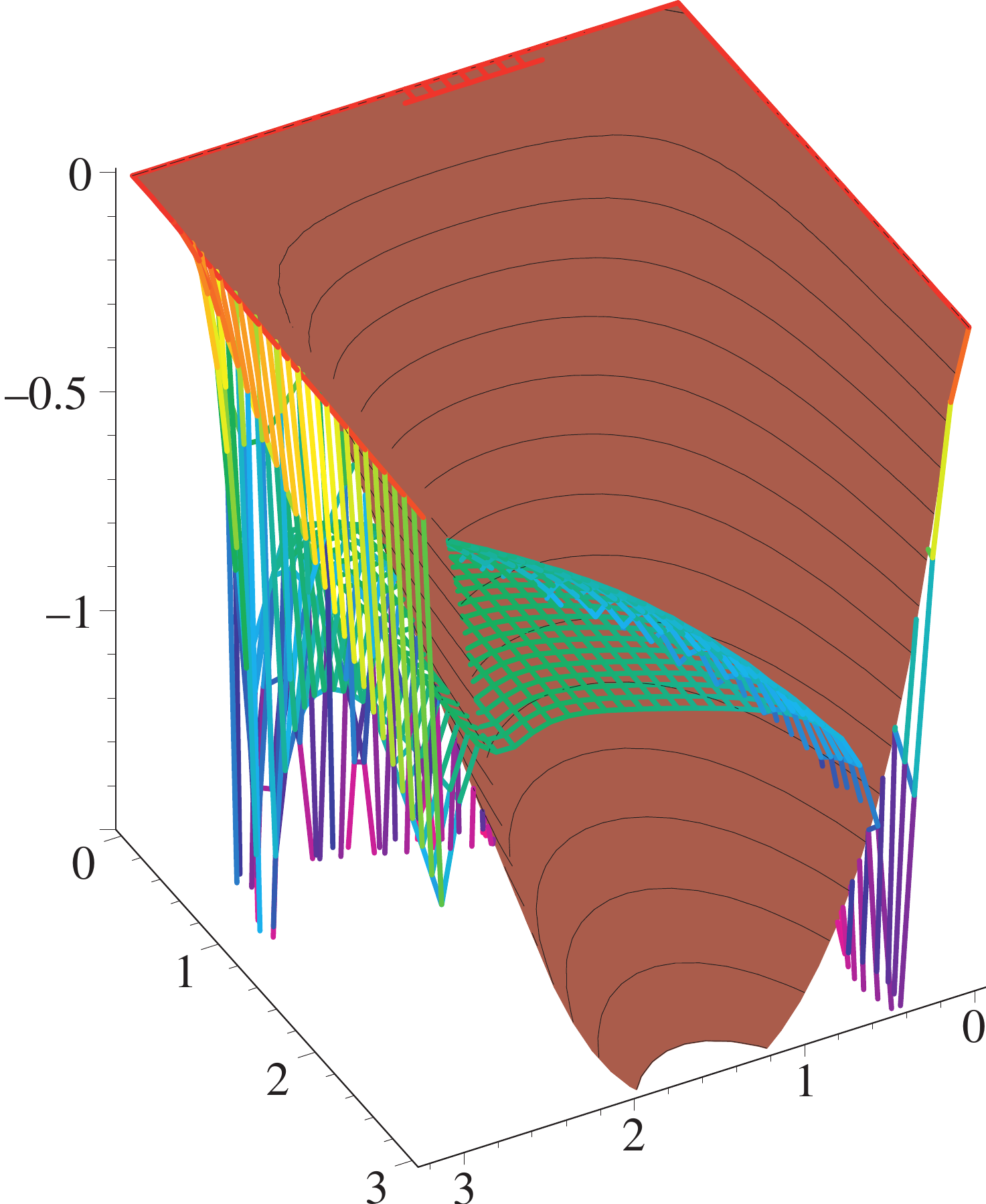}
     \put(-110,105){\rotatebox{0}{\mbox{\bf A}}}
     \put(-124,90){\rotatebox{0}{$ld({\cal V})$}}
     \put(-95,5){\rotatebox{0}{$\theta$}}
     \put(-10,5){\rotatebox{0}{\mbox{$t$}}}
  \end{minipage}
  \hspace{0.035\linewidth}
  \begin{minipage}[b]{0.47\linewidth}
     \includegraphics[width=1\linewidth,height=1\linewidth]{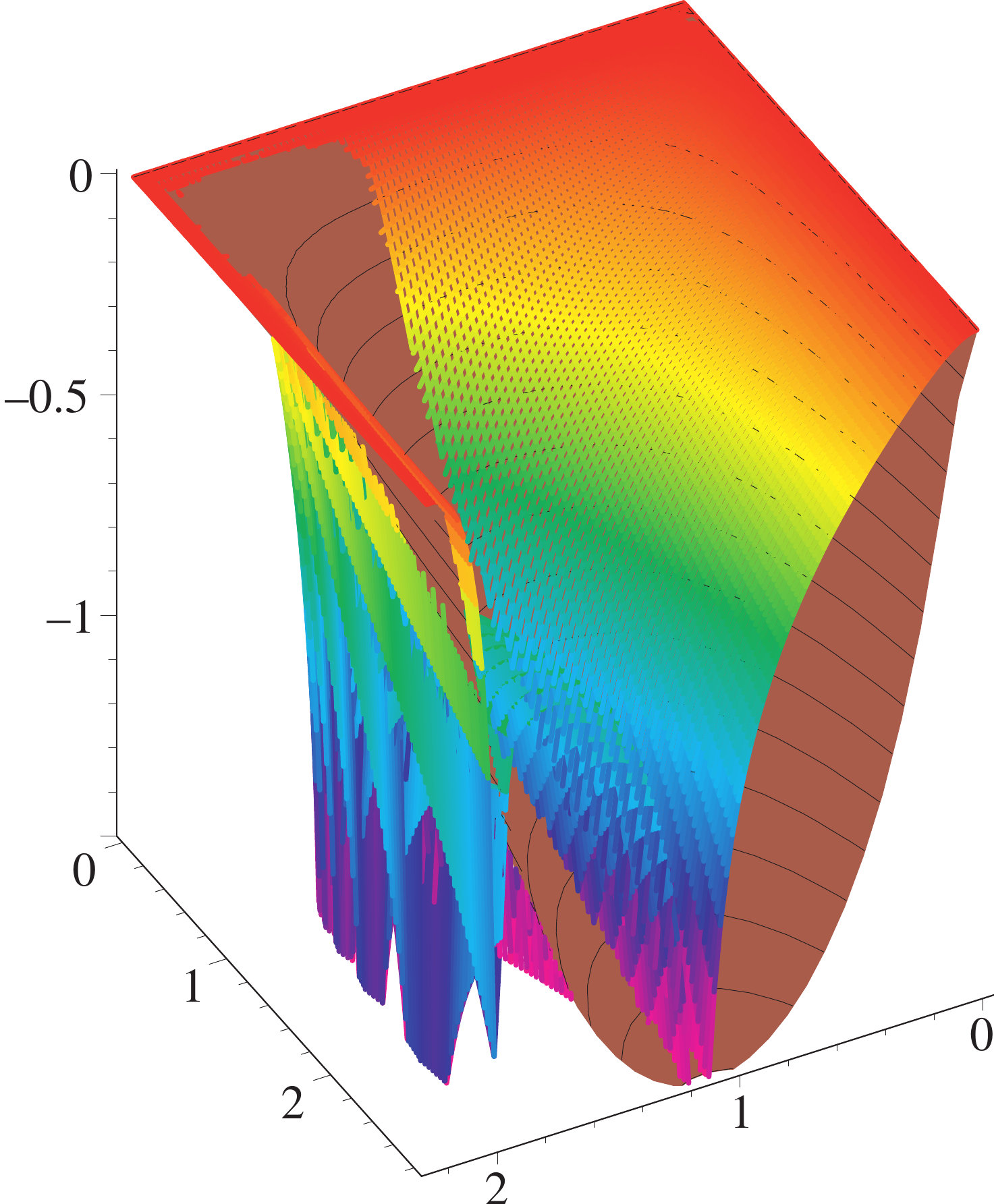}
     \put(-110,105){\rotatebox{0}{\mbox{\bf C}}}
     \put(-124,90){\rotatebox{0}{$ld({\cal V})$}}
     \put(-95,5){\rotatebox{0}{$\theta$}}
     \put(-10,5){\rotatebox{0}{\mbox{$t$}}}
  \end{minipage}
\\
  \begin{minipage}[b]{0.47\linewidth}
     \includegraphics[width=1\linewidth,height=1\linewidth]{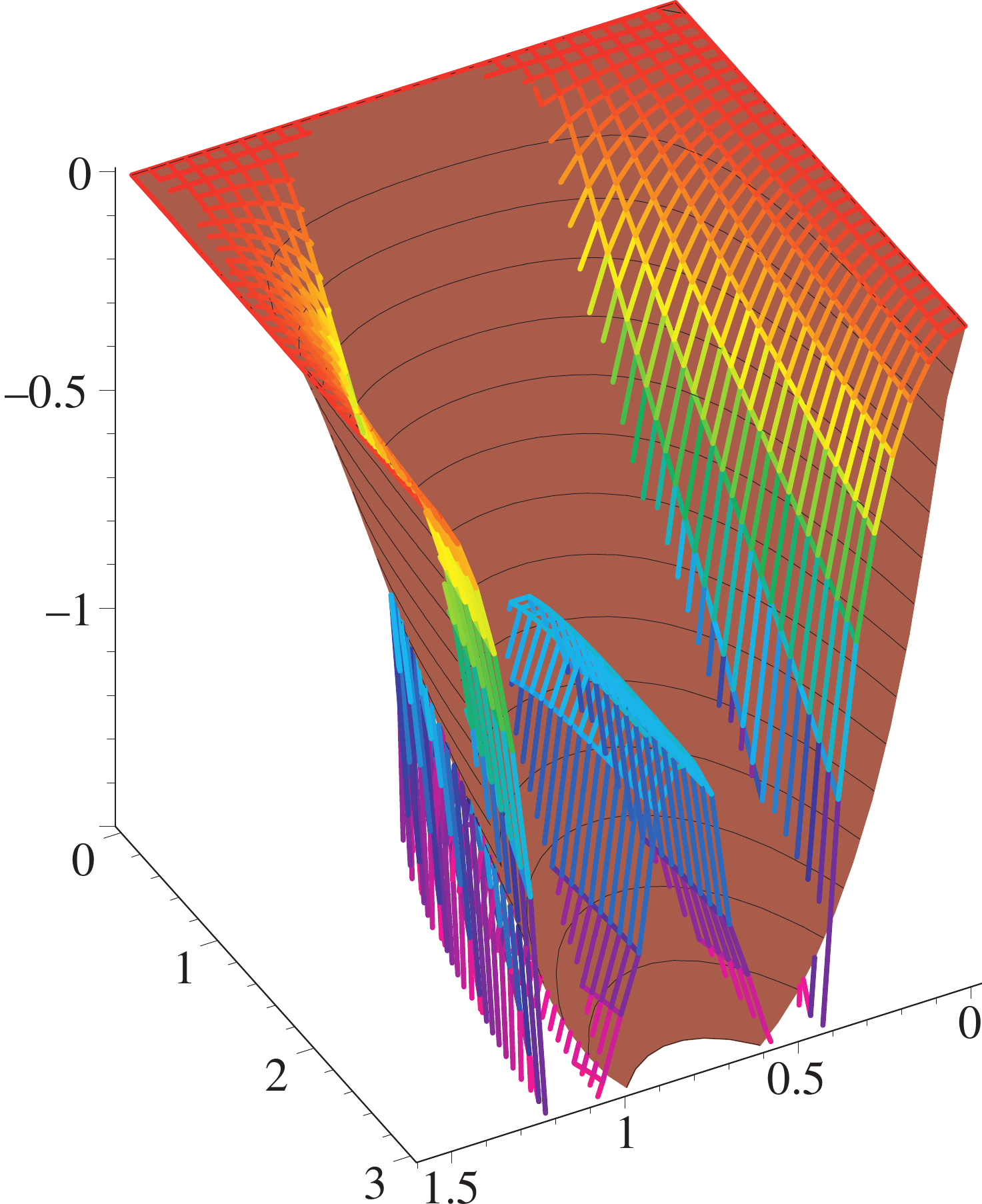}
     \put(-110,105){\rotatebox{0}{\mbox{\bf B}}}
     \put(-124,90){\rotatebox{0}{$ld({\cal V})$}}
     \put(-95,5){\rotatebox{0}{$\theta$}}
     \put(-10,5){\rotatebox{0}{\mbox{$t$}}}
  \end{minipage}
  \hspace{0.035\linewidth}
  \begin{minipage}[b]{0.47\linewidth}
     \includegraphics[width=1\linewidth,height=1\linewidth]{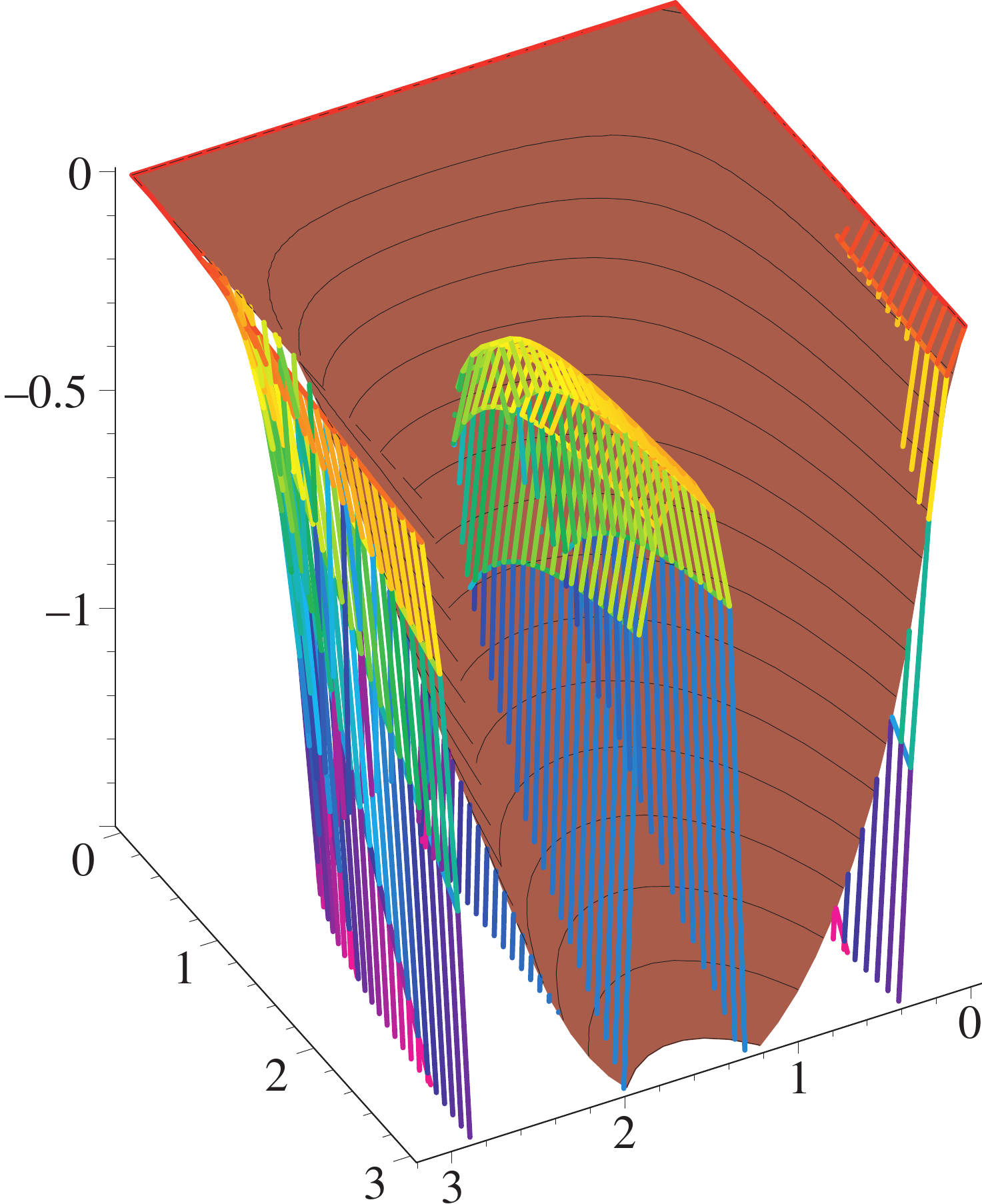}
     \put(-110,105){\rotatebox{0}{\mbox{\bf D}}}
     \put(-124,90){\rotatebox{0}{$ld({\cal V})$}}
     \put(-95,5){\rotatebox{0}{$\theta$}}
     \put(-10,5){\rotatebox{0}{\mbox{$t$}}}
  \end{minipage}
\end{center}
\caption{Same parameters as Fig.~\ref{fig_thermal_states_trapped}
(except for temperature~$\theta$):
  plots of the decadic logarithm $ld({\cal V})$ of the visibility
  $\cal V$ as defined in Eq.~\ref{eq_visibility} as a function of
  temperature and time (colored mesh) together with the benchmark
  approximations~$\cal A$ of Eq.~\ref{eq_visibility_bound} (brown
  contour-sheets).
  This figure shows how with increasing temperature the
  visibility of the interference pattern washes out faster but always
  reforms fully after half a cycle. Moreover, if the mirror gets
  kicked into the momentum superposition state near the turning point
  ($\tau = -T/4$) the interference pattern persists longest, compare
  text and Fig.~\ref{fig_thermal_states_trapped}.}
\label{fig_thermal_states_visibility}
\end{figure}
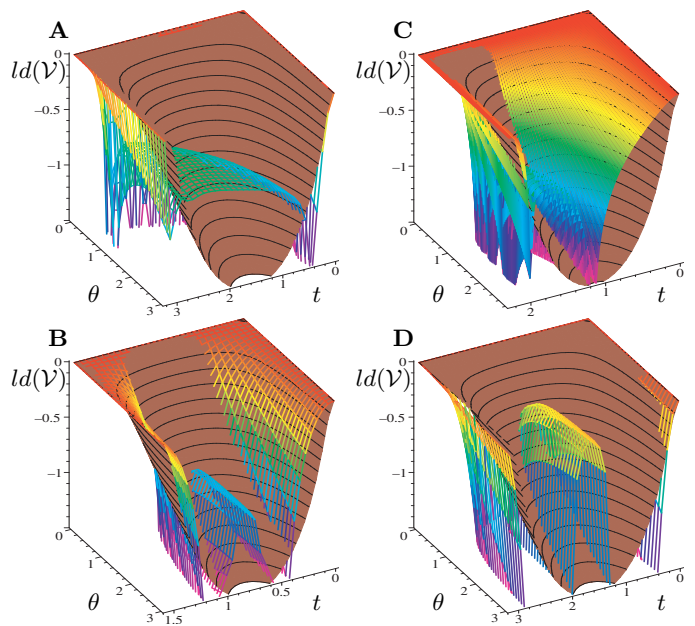

The dephasing times of the interference patterns in evi\-dence in
Figs.~\ref{fig_thermal_states_free}
and~\ref{fig_thermal_states_trapped} can be crudely benchmarked. To
do this we derive a reference approximation~$\cal A$. In the thermal
state the wave functions are incoherently added up and so the
dephasing is due to the summing up of their individual dynamics,
weighted with the Boltzmann-factors. To derive our benchmark
approximation we form a coherent instead of an incoherent sum of
wave functions using the dominant differences in their energy
expressions, namely the harmonic oscillator energies of the initial
stiff trapping potential, in the time evolution phase factors. This
neglects variations due to the transferred kinetic energies but
should be compensated for by the overestimate due to the formation
of a coherent sum.  Moreover, we assume that all wave functions
contribute equally (weight `1') rather than with their true local
weight~$|\Upsilon_n|^2$, this also overestimates the dephasing
effect. This yields an expression containing the `Boltzmann-sums'
$b(\theta,t)=\sum_{n=0}^\infty e^{-(\frac{\Theta_E}{\theta} + i t
\Omega) n} = 1-e^{-(\frac{\Theta_E}{\theta} + i t \Omega)}$. Since
the proba\-bility distribution of the released mirror does not
oscillate at frequency~$\Omega$ but at twice the frequency~$\omega$
of the weak trap, it is released into, our approximation for the
interference pattern visi\-bility adopts this and reads
\begin{eqnarray}
{\cal A}(\theta,t) & = &
\frac{|b(\theta,t \cdot \frac{2 \omega}{\Omega})|^2}{b(\theta,0)^2} \\
& = & \frac{ \left( 1-{e^{-\frac{\Theta_E}{\theta}}} \right)^2 }{1+
e^{-\frac{2 \Theta_E}{\theta}}
-2\,{e^{-\frac{\Theta_E}{\theta}}}\cos \left( 2\,\omega t \right) }
\; . \label{eq_visibility_bound}
\end{eqnarray}
Clearly this approximation is inconsistent; for instance the
probabilities instead of the amplitudes of the Boltzmann-factors are
used, but then again ${\cal A}(\theta/2,t)$ does use the amplitudes,
in short: the approximation $\cal A$ can only serve as a guide to
the eye, compare~Fig.~\ref{fig_thermal_states_visibility}, but it
allows us to underline the fact that dephasing happens on time
scales of the system's evolution $\sim 1/(2\omega)$. This answers
the second question posed in the introduction: Since the approach
using entrained measurements allows for very fast state preparation
and analysis~\cite{Ole_quantumEntrained}, we can hope that, assuming
it can be experimentally implemented, thermal dephasing does not
affect it too badly.

\section{No decoherence without
  dissipation}\label{sec_No_decoherence_Without}


The state~(\ref{eq_final_mixed_state}) shows full interference
contrast at at the time ($t=0$) when it is kicked into the momentum
superposition state~\cite{Ole_quantumEntrained}. The associated
visibility~$\cal V$ of the mirror's interference pattern at the
origin, as a function of time, is given by the expression
\begin{eqnarray}
{\cal V}(t) = \frac{
|\rho(0,t;p_\gamma,\tau,0,k,K_0)-\rho(0,t;p_\gamma,\tau,\pi,k,K_0)
|}{\rho(0,t;p_\gamma,\tau,0,k,K_0)+\rho(0,t;p_\gamma,\tau,\pi,k,K_0)}
\; . \quad  \label{eq_visibility}
\end{eqnarray}
It has this definite form --without the need to maximize or minimize
over the settings of the phase $\phi$-- because all constituent wave
functions have definite parity in the $x$-coordinate.

The visibility degrades over time because of dephasing of the
components of the thermal density
matrix~(\ref{eq_final_mixed_state}). But, if the mirror is released
into a weak harmonic trapping potential with spring constant~$k$
instead of being set free ($k=0$), its wave function will re-form
the same probability distribution (or its position-inverted mirror
image) every half cycle of length~$T/2$; this is illustrated in
Figs.~\ref{fig_pure_states} and~\ref{fig_thermal_states_trapped}.

\subsection{Map between free and trapped
  states}\label{subsec_map_free_trapped}

In theory, the ballistically expanding wave functions can be
determined explicitly, using suitable propagators. In practice, this
turns out to be surprisingly difficult even when using a contemporary
computer algebra system~\cite{Computing_Herts}. An eigenstate of the
initial stiff potential has overlap with many states of a weak
trapping potential it is released into, making the resulting
expressions unwieldy. Instead, a trick is employed here. A mapping has
been used to map between harmonically trapped and free states, for a
detailed discussion see~\cite{Ole_FreeHOSC_arxiv_10}. This way it
became possible to generate Figs.~\ref{fig_thermal_states_trapped}
and~\ref{fig_thermal_states_visibility}.

Aside from being a mathematical trick this mapping also shows that
no decoherence occurs for the case of the free particle, as was
claimed in references~\cite{Ford_PRA01,Ford_PLA01,Ford_AJP02}.
Instead, the wave functions dephase but the mapping back to the
trapped case and further propagation in time shows that the
interference pattern reforms with full contrast. Note that, although
challenging in practice for a widely spread out wave packet, in
principle the mapping corresponds to the experimentalist switching
on the trapping potential~\cite{Ole_FreeHOSC_arxiv_10}. This answers
the first question posed in the introduction: there is no
decoherence without dissipation.

It raises the new question as to when to let the photon impinge on the
mirror to imprint the superposition pattern. The answer to this
question depends on the goals one wants to achieve experimentally.  If
one waits for time~$T/2$ after the release the wave function has
expanded furthest. This stage maybe be advantageous if the large mass
of the mirror makes it impossible to achieve sufficient position
spread early in its evolution. On the other hand, imprinting the
interference pattern at this stage implies that the momentum kick is
transferred to the mirror at a time of largest spatial extension and
subsequently, the mirror will contract again, see
Fig.~\ref{fig_pure_states}~{\bf B}. Unless very large recoil values
are a employed this precludes the formation of separate spatial humps
in the probability distribution which are typically considered to be
the hallmark of so-called ``Schr\"odinger-cat'' spatial superposition
states.

\IgnoreThis{\section{Conclusion}\label{sec_conclusion}}

This work studied effects of thermalized initial states of a
harmonically trapped quantum mirror released into a very weak
potential (or set free)~\cite{Romero-Isart_NJP10,Chang_PNAS10}. It has
shown that decoherence without dissipation, based on thermalized
initial states alone, does not occur as envisaged in
references~\cite{Ford_PRA01,Ford_PLA01,Ford_AJP02}. It also studied
the formation and persistence of interference patterns of the mirror's
wave packet after it is kicked into a superposition state.  It was
shown that dephasing and re-formation of the interference patterns
happens on the time scale of the system's time evolution and should
therefore be detectable when using sufficiently fast detection schemes
such as the `entrainment procedure' sketched in
reference~\cite{Ole_quantumEntrained}.


\bibliography{ThermalCats.bbl}

\begin{thebibliography}{10}%
\makeatletter
\providecommand \@ifxundefined [1]{%
 \ifx #1\undefined \expandafter \@firstoftwo
 \else \expandafter \@secondoftwo
\fi
}%
\providecommand \@ifnum [1]{%
 \ifnum #1\expandafter \@firstoftwo
 \else \expandafter \@secondoftwo
\fi
}%
\providecommand \enquote [1]{``#1''}%
\providecommand \bibnamefont  [1]{#1}%
\providecommand \bibfnamefont [1]{#1}%
\providecommand \citenamefont [1]{#1}%
\providecommand\href[0]{\@sanitize\@href}%
\providecommand\@href[1]{\endgroup\@@startlink{#1}\endgroup\@@href}%
\providecommand\@@href[1]{#1\@@endlink}%
\providecommand \@sanitize [0]{\begingroup\catcode`\&12\catcode`\#12\relax}%
\@ifxundefined \pdfoutput {\@firstoftwo}{%
 \@ifnum{\z@=\pdfoutput}{\@firstoftwo}{\@secondoftwo}%
}{%
 \providecommand\@@startlink[1]{\leavevmode\special{html:<a href="#1">}}%
 \providecommand\@@endlink[0]{\special{html:</a>}}%
}{%
 \providecommand\@@startlink[1]{%
  \leavevmode
  \pdfstartlink
   attr{/Border[0 0 1 ]/H/I/C[0 1 1]}%
   user{/Subtype/Link/A<</Type/Action/S/URI/URI(#1)>>}%
  \relax
 }%
 \providecommand\@@endlink[0]{\pdfendlink}%
}%
\providecommand \url  [0]{\begingroup\@sanitize \@url }%
\providecommand \@url [1]{\endgroup\@href {#1}{\urlprefix}}%
\providecommand \urlprefix [0]{URL }%
\providecommand \Eprint[0]{\href }%
\@ifxundefined \urlstyle {%
  \providecommand \doi [1]{doi:\discretionary{}{}{}#1}%
}{%
  \providecommand \doi [0]{doi:\discretionary{}{}{}\begingroup
  \urlstyle{rm}\Url }%
}%
\providecommand \doibase [0]{http://dx.doi.org/}%
\providecommand \Doi[1]{\href{\doibase#1}}%
\providecommand \bibAnnote [3]{%
  \BibitemShut{#1}%
  \begin{quotation}\noindent
    \textsc{Key:}\ #2\\\textsc{Annotation:}\ #3%
  \end{quotation}%
}%
\providecommand \bibAnnoteFile [2]{%
  \IfFileExists{#2}{\bibAnnote {#1} {#2} {\input{#2}}}{}%
}%
\providecommand \typeout [0]{\immediate \write \m@ne }%
\providecommand \selectlanguage [0]{\@gobble}%
\providecommand \bibinfo [0]{\@secondoftwo}%
\providecommand \bibfield [0]{\@secondoftwo}%
\providecommand \translation [1]{[#1]}%
\providecommand \BibitemOpen[0]{}%
\providecommand \bibitemStop [0]{}%
\providecommand \bibitemNoStop [0]{.\EOS\space}%
\providecommand \EOS [0]{\spacefactor3000\relax}%
\providecommand \BibitemShut [1]{\csname bibitem#1\endcsname}%
\bibitem{Marquardt_Physics.2.40}%
  \BibitemOpen
  \bibfield{author}{%
  \bibinfo {author} {\bibfnamefont{F.}~\bibnamefont{{Marquardt}}}\ and\
  \bibinfo {author} {\bibfnamefont{S.~M.}\ \bibnamefont{{Girvin}}},\ }%
  \bibfield{journal}{%
  \Doi{10.1103/Physics.2.40}{\bibinfo {journal} {Physics}}\ }%
  \textbf{\bibinfo {volume} {2}},\ \bibinfo {eid} {40} (\bibinfo {month} {May}\
  \bibinfo {year} {2009})%
  \bibAnnoteFile{NoStop}{Marquardt_Physics.2.40}%
\bibitem{Ford_PRA01}%
  \BibitemOpen
  \bibfield{author}{%
  \bibinfo {author} {\bibfnamefont{G.~W.}\ \bibnamefont{Ford}}, \bibinfo
  {author} {\bibfnamefont{J.~T.}\ \bibnamefont{Lewis}},\ and\ \bibinfo {author}
  {\bibfnamefont{R.~F.}\ \bibnamefont{O'Connell}},\ }%
  \bibfield{journal}{%
  \Doi{10.1103/PhysRevA.64.032101}{\bibinfo {journal} {Phys. Rev. A}}\ }%
  \textbf{\bibinfo {volume} {64}},\ \bibinfo {pages} {032101} (\bibinfo {month}
  {Aug}\ \bibinfo {year} {2001})%
  \bibAnnoteFile{NoStop}{Ford_PRA01}%
\bibitem{Ford_PLA01}%
  \BibitemOpen
  \bibfield{author}{%
  \bibinfo {author} {\bibfnamefont{G.~W.}\ \bibnamefont{Ford}}\ and\ \bibinfo
  {author} {\bibfnamefont{R.~F.}\ \bibnamefont{O'Connell}},\ }%
  \bibfield{journal}{%
  \bibinfo {journal} {Phys. Lett. A}\ }%
  \textbf{\bibinfo {volume} {286}},\ \bibinfo {pages} {87 } (\bibinfo {year}
  {2001})%
  \bibAnnoteFile{NoStop}{Ford_PLA01}%
\bibitem{Ford_AJP02}%
  \BibitemOpen
  \bibfield{author}{%
  \bibinfo {author} {\bibfnamefont{G.~W.}\ \bibnamefont{Ford}}\ and\ \bibinfo
  {author} {\bibfnamefont{R.~F.}\ \bibnamefont{O'Connell}},\ }%
  \bibfield{journal}{%
  \Doi{10.1119/1.1447540}{\bibinfo {journal} {Am. J. Phys.}}\ }%
  \textbf{\bibinfo {volume} {70}},\ \bibinfo {pages} {319} (\bibinfo {year}
  {2002})%
  \bibAnnoteFile{NoStop}{Ford_AJP02}%
\bibitem{Gobert_PRA04}%
  \BibitemOpen
  \bibfield{author}{%
  \bibinfo {author} {\bibfnamefont{D.}~\bibnamefont{{Gobert}}}, \bibinfo
  {author} {\bibfnamefont{J.}~\bibnamefont{{von Delft}}},\ and\ \bibinfo
  {author} {\bibfnamefont{V.}~\bibnamefont{{Ambegaokar}}},\ }%
  \bibfield{journal}{%
  \Doi{10.1103/PhysRevA.70.026101}{\bibinfo {journal} {\pra}}\ }%
  \textbf{\bibinfo {volume} {70}},\ \bibinfo {pages} {026101} (\bibinfo {month}
  {Aug.}\ \bibinfo {year} {2004}),\
  \Eprint{http://arxiv.org/abs/arXiv:quant-ph/0306019}{arXiv:quant-ph/0306019}%
  \bibAnnoteFile{NoStop}{Gobert_PRA04}%
\bibitem{Ambegaokar_JSP06}%
  \BibitemOpen
  \bibfield{author}{%
  \bibinfo {author} {\bibfnamefont{V.}~\bibnamefont{{Ambegaokar}}},\ }%
  \bibfield{journal}{%
  \Doi{10.1007/s10955-005-8018-6}{\bibinfo {journal} {J. Stat. Phys.}}\ }%
  \textbf{\bibinfo {volume} {125}},\ \bibinfo {pages} {1183} (\bibinfo {month}
  {Dec.}\ \bibinfo {year} {2006}),\
  \Eprint{http://arxiv.org/abs/arXiv:quant-ph/0506087}{arXiv:quant-ph/0506087}%
  \bibAnnoteFile{NoStop}{Ambegaokar_JSP06}%
\bibitem{Zurek_RMP03}%
  \BibitemOpen
  \bibfield{author}{%
  \bibinfo {author} {\bibfnamefont{W.~H.}\ \bibnamefont{Zurek}},\ }%
  \bibfield{journal}{%
  \Doi{10.1103/RevModPhys.75.715}{\bibinfo {journal} {Rev. Mod. Phys.}}\ }%
  \textbf{\bibinfo {volume} {75}},\ \bibinfo {pages} {715} (\bibinfo {month}
  {May}\ \bibinfo {year} {2003})%
  \bibAnnoteFile{NoStop}{Zurek_RMP03}%
\bibitem{Schlosshauer_RMP05}%
  \BibitemOpen
  \bibfield{author}{%
  \bibinfo {author} {\bibfnamefont{M.}~\bibnamefont{Schlosshauer}},\ }%
  \bibfield{journal}{%
  \Doi{10.1103/RevModPhys.76.1267}{\bibinfo {journal} {Rev. Mod. Phys.}}\ }%
  \textbf{\bibinfo {volume} {76}},\ \bibinfo {pages} {1267} (\bibinfo {month}
  {Feb}\ \bibinfo {year} {2005})%
  \bibAnnoteFile{NoStop}{Schlosshauer_RMP05}%
\bibitem{Corbitt_PRL07}%
  \BibitemOpen
  \bibfield{author}{%
  \bibinfo {author} {\bibfnamefont{T.}~\bibnamefont{Corbitt}}, \bibinfo
  {author} {\bibfnamefont{Y.}~\bibnamefont{Chen}}, \bibinfo {author}
  {\bibfnamefont{E.}~\bibnamefont{Innerhofer}}, \bibinfo {author}
  {\bibfnamefont{H.}~\bibnamefont{M\"uller-Ebhardt}}, \bibinfo {author}
  {\bibfnamefont{D.}~\bibnamefont{Ottaway}}, \bibinfo {author}
  {\bibfnamefont{H.}~\bibnamefont{Rehbein}}, \bibinfo {author}
  {\bibfnamefont{D.}~\bibnamefont{Sigg}}, \bibinfo {author}
  {\bibfnamefont{S.}~\bibnamefont{Whitcomb}}, \bibinfo {author}
  {\bibfnamefont{C.}~\bibnamefont{Wipf}},\ and\ \bibinfo {author}
  {\bibfnamefont{N.}~\bibnamefont{Mavalvala}},\ }%
  \bibfield{journal}{%
  \Doi{10.1103/PhysRevLett.98.150802}{\bibinfo {journal} {Phys. Rev. Lett.}}\
  }%
  \textbf{\bibinfo {volume} {98}},\ \bibinfo {pages} {150802} (\bibinfo {month}
  {Apr}\ \bibinfo {year} {2007})%
  \bibAnnoteFile{NoStop}{Corbitt_PRL07}%
\bibitem{Ole_quantumEntrained}%
  \BibitemOpen
  \bibfield{author}{%
  \bibinfo {author} {\bibfnamefont{O.}~\bibnamefont{Steuernagel}}}%
   (\bibinfo {year} {2011}),\
  \Eprint{http://arxiv.org/abs/1106.0202}{arXiv:1106.0202}%
  \bibAnnoteFile{NoStop}{Ole_quantumEntrained}%
\bibitem{Ole_PRA95}%
  \BibitemOpen
  \bibfield{author}{%
  \bibinfo {author} {\bibfnamefont{O.}~\bibnamefont{Steuernagel}}\ and\
  \bibinfo {author} {\bibfnamefont{H.}~\bibnamefont{Paul}},\ }%
  \bibfield{journal}{%
  \Doi{10.1103/PhysRevA.52.R905}{\bibinfo {journal} {Phys. Rev. A}}\ }%
  \textbf{\bibinfo {volume} {52}},\ \bibinfo {pages} {R905} (\bibinfo {month}
  {Aug}\ \bibinfo {year} {1995})%
  \bibAnnoteFile{NoStop}{Ole_PRA95}%
\bibitem{Thompson_NAT08}%
  \BibitemOpen
  \bibfield{author}{%
  \bibinfo {author} {\bibfnamefont{J.~D.}\ \bibnamefont{{Thompson}}}, \bibinfo
  {author} {\bibfnamefont{B.~M.}\ \bibnamefont{{Zwickl}}}, \bibinfo {author}
  {\bibfnamefont{A.~M.}\ \bibnamefont{{Jayich}}}, \bibinfo {author}
  {\bibfnamefont{F.}~\bibnamefont{{Marquardt}}}, \bibinfo {author}
  {\bibfnamefont{S.~M.}\ \bibnamefont{{Girvin}}},\ and\ \bibinfo {author}
  {\bibfnamefont{J.~G.~E.}\ \bibnamefont{{Harris}}},\ }%
  \bibfield{journal}{%
  \Doi{10.1038/nature06715}{\bibinfo {journal} {\nat}}\ }%
  \textbf{\bibinfo {volume} {452}},\ \bibinfo {pages} {72} (\bibinfo {month}
  {Mar.}\ \bibinfo {year} {2008}),\
  \Eprint{http://arxiv.org/abs/0707.1724}{arXiv:0707.1724}%
  \bibAnnoteFile{NoStop}{Thompson_NAT08}%
\bibitem{Groeblacher_NATPhys09}%
  \BibitemOpen
  \bibfield{author}{%
  \bibinfo {author} {\bibfnamefont{S.}~\bibnamefont{{Gr{\"o}blacher}}},
  \bibinfo {author} {\bibfnamefont{J.~B.}\ \bibnamefont{{Hertzberg}}}, \bibinfo
  {author} {\bibfnamefont{M.~R.}\ \bibnamefont{{Vanner}}}, \bibinfo {author}
  {\bibfnamefont{G.~D.}\ \bibnamefont{{Cole}}}, \bibinfo {author}
  {\bibfnamefont{S.}~\bibnamefont{{Gigan}}}, \bibinfo {author}
  {\bibfnamefont{K.~C.}\ \bibnamefont{{Schwab}}},\ and\ \bibinfo {author}
  {\bibfnamefont{M.}~\bibnamefont{{Aspelmeyer}}},\ }%
  \bibfield{journal}{%
  \Doi{10.1038/nphys1301}{\bibinfo {journal} {Nature Phys.}}\ }%
  \textbf{\bibinfo {volume} {5}},\ \bibinfo {pages} {485} (\bibinfo {month}
  {Jul.}\ \bibinfo {year} {2009}),\
  \Eprint{http://arxiv.org/abs/0901.1801}{arXiv:0901.1801}%
  \bibAnnoteFile{NoStop}{Groeblacher_NATPhys09}%
\bibitem{Ole_FreeHOSC_arxiv_10}%
  \BibitemOpen
  \bibfield{author}{%
  \bibinfo {author} {\bibfnamefont{O.}~\bibnamefont{Steuernagel}}}%
   (\bibinfo {year} {2010}),\
  \Eprint{http://arxiv.org/abs/1008.3929}{arXiv:1008.3929}%
  \bibAnnoteFile{NoStop}{Ole_FreeHOSC_arxiv_10}%
\bibitem{HongOuMandel87}%
  \BibitemOpen
  \bibfield{author}{%
  \bibinfo {author} {\bibfnamefont{C.~K.}\ \bibnamefont{Hong}}, \bibinfo
  {author} {\bibfnamefont{Z.~Y.}\ \bibnamefont{Ou}},\ and\ \bibinfo {author}
  {\bibfnamefont{L.}~\bibnamefont{Mandel}},\ }%
  \bibfield{journal}{%
  \Doi{10.1103/PhysRevLett.59.2044}{\bibinfo {journal} {Phys. Rev. Lett.}}\ }%
  \textbf{\bibinfo {volume} {59}},\ \bibinfo {pages} {2044} (\bibinfo {month}
  {Nov}\ \bibinfo {year} {1987})%
  \bibAnnoteFile{NoStop}{HongOuMandel87}%
\bibitem{Computing_Herts}%
  \BibitemOpen
  \bibinfo {note} {{This work made use of the high-performance computing
  facility at the University of Hertfordshire's Science and Technology Research
  Institute.}}%
  \bibAnnoteFile{Stop}{Computing_Herts}%
\bibitem{Romero-Isart_NJP10}%
  \BibitemOpen
  \bibfield{author}{%
  \bibinfo {author} {\bibfnamefont{O.}~\bibnamefont{{Romero-Isart}}}, \bibinfo
  {author} {\bibfnamefont{M.~L.}\ \bibnamefont{{Juan}}}, \bibinfo {author}
  {\bibfnamefont{R.}~\bibnamefont{{Quidant}}},\ and\ \bibinfo {author}
  {\bibfnamefont{J.~I.}\ \bibnamefont{{Cirac}}},\ }%
  \bibfield{journal}{%
  \Doi{10.1088/1367-2630/12/3/033015}{\bibinfo {journal} {New J. Phys.}}\ }%
  \textbf{\bibinfo {volume} {12}},\ \bibinfo {pages} {033015} (\bibinfo {month}
  {Mar.}\ \bibinfo {year} {2010}),\
  \Eprint{http://arxiv.org/abs/0909.1469}{arXiv:0909.1469 [quant-ph]}%
  \bibAnnoteFile{NoStop}{Romero-Isart_NJP10}%
\bibitem{Chang_PNAS10}%
  \BibitemOpen
  \bibfield{author}{%
  \bibinfo {author} {\bibfnamefont{D.~E.}\ \bibnamefont{Chang}}, \bibinfo
  {author} {\bibfnamefont{C.~A.}\ \bibnamefont{Regal}}, \bibinfo {author}
  {\bibfnamefont{S.~B.}\ \bibnamefont{Papp}}, \bibinfo {author}
  {\bibfnamefont{D.~J.}\ \bibnamefont{Wilson}}, \bibinfo {author}
  {\bibfnamefont{J.}~\bibnamefont{Ye}}, \bibinfo {author}
  {\bibfnamefont{O.}~\bibnamefont{Painter}}, \bibinfo {author}
  {\bibfnamefont{H.~J.}\ \bibnamefont{Kimble}},\ and\ \bibinfo {author}
  {\bibfnamefont{P.}~\bibnamefont{Zoller}},\ }%
  \bibfield{journal}{%
  \Doi{10.1073/pnas.0912969107}{\bibinfo {journal} {Proc. Nat. Acad. Sci.}}\ }%
  \textbf{\bibinfo {volume} {107}},\ \bibinfo {pages} {1005} (\bibinfo {year}
  {2010})%
  \bibAnnoteFile{NoStop}{Chang_PNAS10}%
\end{thebibliography}%

\end{document}